\documentclass{aa}

\usepackage{natbib}
\usepackage{graphicx}
\usepackage{txfonts,textcomp}

\begin{document}

\title{Carbon accretion and desorption by interstellar polycyclic aromatic hydrocarbons}
\titlerunning{PAH carbon return to the ISM}

\author{A. Omont\inst{1}
\and H.\ F.\ Bettinger\inst{2}
}
\institute{
Sorbonne Universit\'e, UPMC Universit\'e Paris 6 and CNRS, UMR 7095, Institut d'Astrophysique de Paris, France\\
\email{omont@iap.fr}
\and  Institut für Organische Chemie, Universität Tübingen, Auf der Morgenstelle 18, 72076 Tübingen, Germany\\
\email{holger.bettinger@uni-tuebingen.de}
}

\abstract{Two key questions of the chemistry of polycyclic aromatic hydrocarbons (PAHs) in the interstellar medium (ISM) are addressed: (i) the way carbon is returned from PAHs to the interstellar gas after the very efficient accretion of C$^+$ ions onto PAHs, and (ii) the PAH contribution to the high abundance of small carbon molecules observed in UV-irradiated regions. They are addressed based on the structure and stability of the various isomers of the complexes formed by PAHs and their cations with atomic carbon. Carbon  complexes with coronene are studied by B3LYP/6-311+G** calculations,  in order to determine the behaviour of C$^+$ and C complexes with larger pericondensed interstellar PAHs, which are thought to be dominant in the ISM. The most stable forms of the [C-coronene]$^+$ cation include  7C and 4C rings, C$^+$ insertion into a CH bond, and a 5C ring with a short exocyclic cumulene chain, and similarly for neutral [C-coronene]. The subsequent evolution of similar complexes with pericondensed PAHs, in diffuse clouds, is discussed under the action of interstellar UV photons and H atoms as a function of the PAH size. Despite the complexity of this processing,
 it seems probable that, for small PAHs, these complexes efficiently lose a C$_2$H$_2$ molecule from repeated photodissociations.  However, this conclusion needs to be confirmed by the identification of reaction paths and the computation of activation energies. The case of the evolution of larger [C-PAH] complexes is less clear. The processing may explain the observed balance between C$^+$ and PAHs, at least in the diffuse ISM. The formation of C$_2$H$_2$ from PAH catalysis is a key input for the chemistry of small carbon molecules in diffuse clouds. C$^+$ accretion might frequently form stable PAHs that contain a peripheral pentagonal ring and form a significant fraction of interstellar PAHs.} 


\maketitle 

\section{Introduction}
\begin{figure}
        \begin{center}
                \includegraphics[scale=0.38, angle=0]{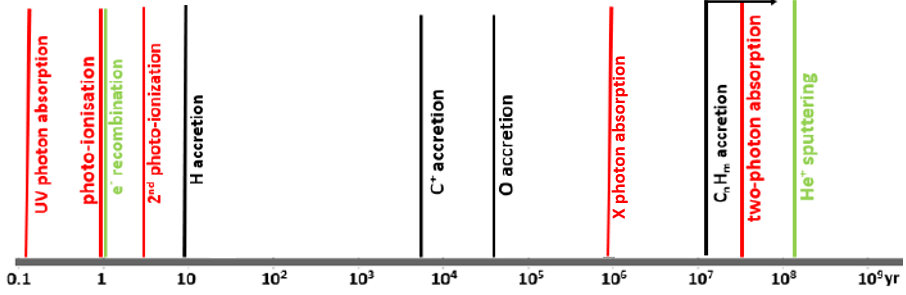}
                \caption{(Adapted from Omont \& Bettinger 2021). Estimated orders of magnitude of reaction rates for a typical PAH/PAH$^+$ of 50 C atoms in a diffuse interstellar cloud with n$_{\rm HT}$\,=\,50\,cm$^{-3}$, T\,=\,50-100\,K and UV intensity G$_0$\,=\,3\,Habing. References for these rates are given in Table A.1 of Omont \& Bettinger (2021), except for the rate of H accretion that has been slightly increased to match the adopted value of $\sim$1.4 10$^{-10}$ cm$^{3}$ s$^{-1}$ by Andrews et al.\ (2016) for any H--PAH$^+$ accretion, assuming equal fractions of neutral and cation PAHs. While these rates have been estimated for typical  PAHs, they should also apply to [C-PAH] complexes without much change, except for H-accretion that might proceed slightly  faster (Appendix \ref{app:C.3Hydrogenation}).}
                \label{fig:6A1-rates}
        \end{center}
\end{figure}

The 3--12\,$\mu$m infrared spectrum of galaxies and many Galactic sources, including photon-dominated regions (PDRs) and the diffuse interstellar medium (ISM), exhibits strong emission bands characteristic of C-H and C-C vibrations of polycyclic aromatic compounds. The band intensity ratios, especially the strong intensity of the 3.3\,$\mu$m band, point   to high excitation temperature (up to $\sim$1000\,K). This implies transient sporadic heating by a single UV photon of aromatic nanoparticles with a number of carbon atoms N$_{\rm C}$\,$\la$\,100--150, 
that are mainly polycyclic aromatic hydrocarbons (PAHs) and their derivatives (Leger \& Puget 1984; Allamandola et al.\ 1985; Puget \& Leger 1989; Tielens 2008; Knight et al.\ 2021).
They include a substantial fraction ($\sim$5--10\%) of the total abundance of carbon in the diffuse ISM. By extending
 the size distribution of interstellar carbon dust grains, their total surface is comparable to that of dust grains (Omont 1986). Therefore, they dominate the accretion of gaseous particles on interstellar carbonaceous grains and nanoparticles so that they play an important role in the chemistry of PDRs and the diffuse ISM (Omont 1986, Lepp \& Dalgarno 1988; Lepp et al.\ 1988; Tielens 2008, 2013).
Depending on the UV intensity and the gas density, their main charge states are generally neutral or singly charged cations, with frequent changes of charge  (see e.g.\  Bakes \& Tielens 1994; Bakes et al.\ 2001, and Fig.\ \ref{fig:6A1-rates}). However,  anions or doubly charged cations may be significant in regions of low or very high UV intensity, respectively.

In typical conditions of the diffuse ISM (density of H nuclei n$_{\rm HT}$ = 30–100 cm$^{-3}$, temperature T = 50–100 K, and UV intensity G$_0$ = 1–10 Habing), there is a very broad hierarchy in the rates of the different chemical processes affecting a typical PAH, as visualized in Fig.\ \ref{fig:6A1-rates} (see Table A.1 and Fig.\ 1 of Omont \& Bettinger 2021). They are dominated by photochemistry since the rate of absorption of UV photons is two orders of magnitude faster than the accretion of H atoms, and more than four or five orders of magnitude faster than the accretion of C$^+$ or O. More frequent than photoionization, the dominant process is photodissociation, where the loss of H is easier than that of small carbonaceous fragments. Its complexity is well documented by many studies (see Sect. \ref{sec:4.1summphotdiss}).

The  other main chemical processes are reactions with the most abundant constituents of the  diffuse interstellar gas,  H, C$^+$, and O, resulting mainly in their accretion onto PAHs. Hydrogen reactions are by far the most frequent; they  produce rehydrogenation of dehydrogenated PAHs, superhydrogenation of PAHs (Appendix \ref{app:BsuperH}), and H$_2$ formation (e.g.\  Andrews et al.\ 2016; Castellanos et al.\ 2018a,b). 

C$^+$ accretion is known to be fairly efficient
 (Canosa et al.\ 1995; see Sect. \ref{sec:2.2CpPAH}). If it was not compensated for by efficient processes for carbon desorption from PAHs, it would deplete interstellar C$^+$ within a few  10$^6$\,yr (Eq. \ref{eq:D3tauC}), jeopardizing the actual existence of PAHs as carriers of the 3 to 12\,$\mu$m infrared bands (e.g.\ Thaddeus 1995; Le Page et al.\ 2001, 2003; Tielens 2008; Kwok 2021, and references therein).  However, the way carbon is returned from PAHs to the gas phase of the ISM  does not seem to have been elucidated in detail. The purpose of this paper is to further assess the nature of the dominant interstellar   
 [C-PAH] complexes  resulting from C$^+$ accretion and the way their photoprocessing returns carbon to the gas in the diffuse ISM.  We conclude that (i) for the smallest PAHs, at least  one-half the size distribution, carbon is  probably mainly quickly returned as  C$_2$H$_2$, in particular from aliphatic PAH derivatives (similarly to Marciniak et al.\ 2021), and (ii) for the largest PAHs, the balance is uncertain between rapid C$_2$H$_2$ loss or PAH growth followed by eventual fragmentation.
This should contribute to the high abundances of C$_2$H$_x$ molecules observed in the diffuse ISM (Liszt et al.\ 2018) and PDRs (Pety et al.\ 2005; Guzm{\'a}n et al. 2015).

\begin{figure*}[htbp]
         \begin{center}
\includegraphics[scale=1.1,angle=0]{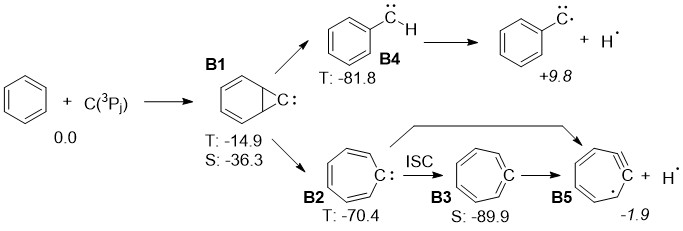}
 \caption{Simplified reaction scheme of the C($^3$P$_{\rm j}$) + benzene system, as computed by Bettinger et al. (2000). The energies (kcal/mol) for spin-triplet (T) and spin-singlet (S) species were obtained at B3LYP/6-311+ZPVE(B3LYP/6-31G*) (normal print), while the energies of C$_7$H$_5$ radicals were computed at G2(B3LYP/MP2) (in italics).}
 \label{fig:1-figure1}
     \end{center}
 \end{figure*}

The paper is organized as follows. Section \ref{sec:2basic} is devoted to recalling the known basic features of the reaction of  C and C$^+$ with benzene and PAHs. Section \ref{sec:3CCpPAH}
 addresses the relative stability of [C-PAH]$^+$ and [C-PAH] complexes by DFT calculations of the model case of various [C-coronene]$^+$ and [C-coronene] complexes. In Sect. \ref{sec:4smallCPAH} we analyse the photodissociation of small [C-PAH] complexes in the diffuse ISM and the way they might quickly return C$_2$H$_2$ to the gas. 
In Sect. \ref{sec:5largePAHs}  we extend the discussion to larger PAHs and multi-C$^+$ accretions. 
In Sect. \ref{sec:7CO} we consider the minor effect of reactions of atomic O with PAH$^+$ cations, probably resulting in PAH sputtering through CO ejection.   
In Sect. \ref{sec:8astrophysics} we examine the possible astrophysical consequences for the diffuse ISM, as regards the observed balance between gas C$^+$ and PAHs, and the chemistry of small carbon molecules from the injection of C$_2$H$_2$ molecules. Finally, in Sect. \ref{sec:9conclusion} we present  our main conclusions from this study.

\section{Basic features of the reaction of  C and C$^+$ with PAHs}
\label{sec:2basic}

\begin{figure*}[htbp]
         \begin{center}
\includegraphics[scale=1.1, angle=0]{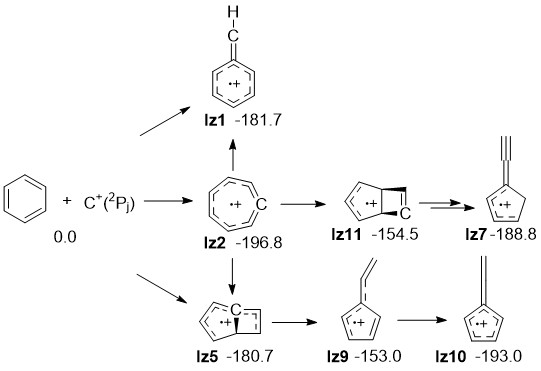}
 \caption{Simplified reaction scheme of the  C$^+$($^2$P$_{\rm j}$) + benzene system, as computed by Izadi et al. (2021). The energies (kcal/mol) were obtained at PBE0-D3(BJ)/aug-cc-pVTZ + ZPVE(PBE0-D3(BJ)/aug-cc-pVTZ). The depiction of the structures and their numbering follow that of Izadi et al.\ (2021).}
 \label{fig:2-figure2}
     \end{center}
 \end{figure*}

\subsection{Reaction of atomic C and C$^+$ with benzene and with small PAHs}
\label{sec:2.1benzene}
        The reaction of C($^3$P$_{\rm j}$) with benzene was investigated experimentally and computationally (Kaiser et al.\ 1999; Bettinger et al.\ 2000; Hahndorf et al.\ 
        2002; Krasnokutski \& Huisken 2014; Izadi et al.\ 2021). Crossed-beam experiments at various collision energies combined with ab initio and Rice–Ramsperger–Kassel–Marcus (RRKM) calculations 
        suggest a barrierless reaction in the entrance and exit channels and the formation of a seven-membered ring C$_7$H$_5$ radical based on the potential energy surface of 
        the C$_7$H$_6$ system (Fig.\ \ref{fig:1-figure1}; Kaiser et al. 1999, Bettinger et al. 2000, Hahndorf et al. 2002). The carbon atom adds to the benzene edge (B1) to a bicyclic 
        intermediate. Opening   the three-membered ring results in a seven-membered ring (B2). In the gas phase the high reaction energy gained can be channelled into 
        breaking   a CH bond to give the observed C$_7$H$_5$ seven-ring product (B5). In ultra-cold He droplets the reaction gives the seven-membered C$_7$H$_6$ isomer 
        cycloheptatetraene (B3) after intersystem crossing (ISC) as the reaction energy is dissipated to the surrounding matrix (Krasnokutski \& Huisken 2014).

Similar results were derived by Izadi et al.\ (2021), who extended the calculations to the C$^+$($^2$P$_{\rm j}$) + C$_6$H$_6$ system, as shown in Fig.\ \ref{fig:2-figure2}. The possible reaction 
scheme identified is more complex than that for the reaction with neutral C($^3$P$_{\rm j}$) and is summarized briefly here. Benzene reacts without barrier to three possible 
metastable complexes,   
to give  either a seven-membered ring (Iz2) or  a bicyclic structure with a pentagonal ring (Iz5), 
or insertion into a C-H bond to give [C$_6$H$_5$-CH]$^+$ that maintains a hexagonal ring (Iz1). We note that   interconversion between the three major products is also conceivable, 
due to the very high exothermicity of the reactions. The most stable product of [C-benzene]$^+$ is again a monocyclic seven-membered ring (Iz2). However, 
five-membered rings with 2C chain adducts, Iz10 and Iz7, have energies that are higher by only a few kcal/mol, while the CH insertion to give [C$_6$H$_5$CH]$^+$, Iz1, has an energy 
higher by 15 kcal/mol. The measurements by Smith \& Futrell (1978, see also Smith \& DeCorpo 1976)
 show that 14\% of the reaction C$^+$ + C$_6$H$_6$ yield dissociative charge exchange with the possible formation of C$_2$H$_3$ or C$_2$H$_2$.

Laboratory and computational studies by Krasnokutski et al.\ (2017) suggest that all 
catacondensed PAHs also react without barrier with atomic 
carbon, while coronene and large compact pericondensed PAHs could be more inert  towards such a reaction.

\subsection{Accretion of interstellar C$^+$ onto PAHs}
\label{sec:2.2CpPAH}

The C$^+$ ion is by far the dominant form of gaseous carbon in regions of the ISM where strong UV radiation is present, such as diffuse clouds and PDRs. It should efficiently accrete onto PAHs, as inferred from the measurement of the reaction of C$^+$ + anthracene (C$_{14}$H$_{10}$) by Canosa et al.\ (1995).
In addition to pure charge exchange occurring in 60\% of the cases, 40\% of the reaction yield C$_{15}$H$_{9}^+$ (i.e. carbon accretion with H-loss dissociative charge exchange). From this measurement, as summarized in Appendix A.3 of Omont \&  Bettinger (2021), one may estimate that the rate of accretion of C$^+$ onto typical neutral PAHs in the diffuse ISM is at least 2$\times$10$^{-9}$\,cm$^3$s$^{-1}$. This  yields about 2$\times$10$^{-4}$\,yr$^{-1}$ for the average rate of C$^+$  accretion by a PAH in the standard diffuse ISM, assuming equal fractions of neutral and cation PAHs (Fig.\ \ref{fig:6A1-rates}). Furthermore,  one can estimate that, in the diffuse ISM, the characteristic time for the depletion of gaseous carbon  by accretion of  C$^+$  onto all PAHs is a few 10$^6$\,yr (Eq.\ \ref{eq:D3tauC}). This confirms that this time is much shorter than all characteristic evolution times of diffuse clouds and their PAHs, which are estimated to be $\ga$\,10$^8$\,yr (Micelotta et al.\ 2010a,b, 2011).

\section{Energies of various isomers of [C-PAH] and [C-PAH$^+$] complexes}  
\label{sec:3CCpPAH}

In order to determine  the most probable  structures of [C-PAH]$^+$ and [C-PAH] complexes resulting from the accretion of a C$^+$ ion onto interstellar PAHs, we  calculated the energies of various isomers of carbon complexes of a relatively small PAH, coronene (C$_{24}$H$_{12}$), that is thought to be representative of the structure of pericondensed PAHs, which should be dominant in interstellar conditions (e.g. Roelfsema et al.\ 1996; Tielens 2013; Andrews et al.\ 2015; Bouwman et al.\ 2019; Chown et al.\ 2024).

\subsection{Method of computation of [C-coronene]$^+$ and [C-coronene] complexes} 
\label{sec:3.1method}

The geometries of stationary points were fully optimized using the B3LYP (Becke et al.\ 1988) hybrid functional in conjunction with the 6-311+G** basis set. Harmonic vibrational frequencies were computed analytically to allow correction of electronic energies for zero-point vibrational energies (ZPVEs). The spin-restricted formalism was used for singlet electronic states, while the spin-unrestricted formalism was employed in the case of doublet and triplet electronic spin states. 

 Previous work on the reaction of benzene and C($^3$P) has shown that the B3LYP/6-311+G** reaction energy of -2.7 kcal/mol compares very well with experimental values (-3.7 ± 1.1 kcal/mol) and the accurate composed methods G2(MP2)//B3LYP (-1.9 kcal/mol) and CBS-Q (-6.2 kcal/mol; Bettinger et al. 2000). To judge the reliability of the B3LYP/6-311+G** method further, we   computed the energies of the products of the reaction of benzene with C($^3$P) and C$^+$($^2$P) depicted in Figs. 2 and 3 using the G3(MP2)//B3LYP (Baboul et al.\ 1999) and CBS-QB3 (Montgomery et al.\ 1999, 2000) model chemistries that attempt to provide highly reliable thermochemistry data. The mean absolute deviation (MAD) of the B3LYP data for the benzene + C($^3$P) reaction compared to G3(MP2)//B3LYP is 3.1 kcal/mol, while it is only 2.1 kcal/mol compared to CBS-QB3. The MAD for the benzene + C$^+$($^2$P) reaction is much larger, 12.6 kcal/mol and 11.0 kcal/mol with respect to G3(MP2)//B3LYP and CBS-QB3, respectively. We note that the PBE0-D3BJ/aug-cc-pVTZ model  employed by Izadi et al.\ (2021) has an even  larger MAD of 20.6 kcal/mol (G3(MP2)//B3LYP) and 19.1 kcal/mol (CBS-QB3). All computations were performed with Gaussian 16 (Frisch et al.\ 2016.). 

\subsection{Energies of cationic [C-coronene]$^+$ complexes}
\label{sec:3.2Cpcoronene}

\begin{figure*}[htbp]
         \begin{center}
\includegraphics[scale=1.1, angle=0]{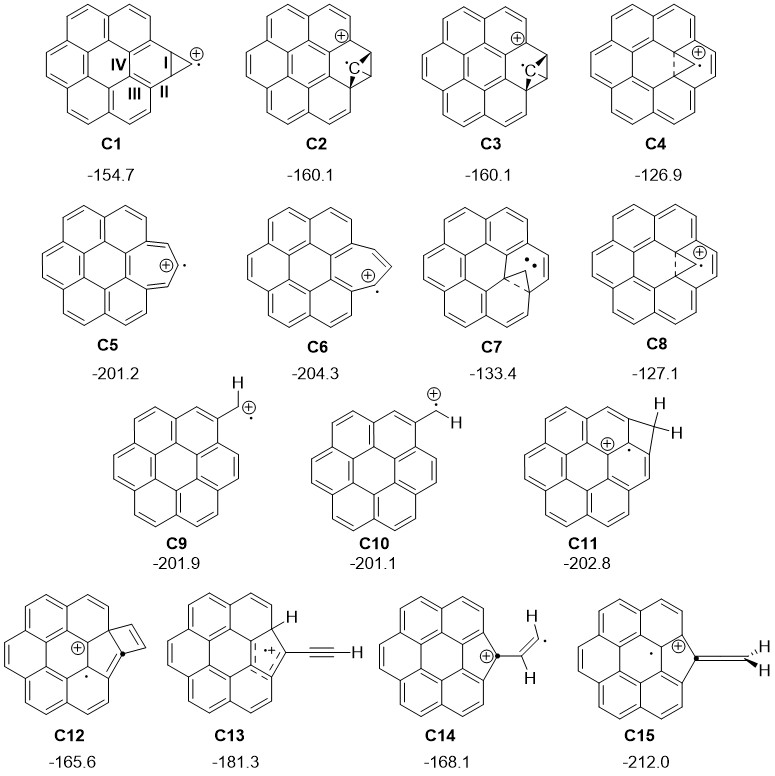}
 \caption{Energies (kcal/mol) of various isomers of cation [C-coronene]$^+$ complexes, C$_{25}$H$_{12}^+$, as computed at B3LYP/6-311+G** + ZPVE (see Sect. \ref{sec:3.1method}). The four symmetry-unique bonds of parent coronene are labelled  I-IV. Structure C1 results from the interaction of C$^+$ with bond I. We note that an attack of bonds II and III by C$^+$ results in identical structures C2 and C3.}
 \label{fig:3-figure3}
     \end{center}
 \end{figure*}

Coronene is of  high symmetry ($D_{6h}$) and has four symmetry-unique CC bonds. We investigated the products from reaction of C$^+$($^2$P$_{\rm J}$) with each of these bonds by 
bridging structures  and insertion (C5-C8) to give seven-membered rings (Fig.\ \ref{fig:3-figure3}). Single and double CH insertion products and structures with 
pentagonal rings, identified as particularly stable for [C-benzene]$^+$ by Izadi et al.\ (2021; see also Marciniak et al.\ 2021), were also investigated. Formation of the bridging complexes is highly 
exothermic, similar to the case of [C-benzene]$^+$.  
These initial complexes are expected to readily form 
stable compounds by CC insertion as concluded for [C-benzene]$^+$. The most stable CC insertion product, C$6$, is 
obtained by C$^+$ insertion into an outer 6C ring to form a 7C ring. Insertion into internal CC bonds (C7 and C8) is much less favourable as the structural 
distortion creates strain on the coronene `backbone'.
 However, as for [C-benzene]$^+$ (Fig.\ \ref{fig:2-figure2}), other structures have comparable energies, such as those obtained 
by C$^+$ insertion into a CH bond (C9-C10). A double CH insertion product  that features a four-membered ring (C11), which is without equivalent for [C-benzene]$^+$, 
is of similar energy to CH insertion products (Fig.\ \ref{fig:3-figure3}). More stable than the 4C ring is a structure with a pentagonal ring 
and an exocyclic cumulene CCH$_2$ (C15), similar  to Iz10 of Izadi et al.\ (2021). This is the most stable form of [C-coronene]$^+$. Its energy is about 8\,kcal/mol lower than the second most stable isomer C6 with a 7C ring (Fig.\ \ref{fig:3-figure3}). The similar structure of [C-benzene]$^+$ is less stable than 
the 7C ring by 3.8 kcal/mol (see Fig.\ \ref{fig:2-figure2}). Other pentagonal structures (C12, C13, C14), similar  to quite  low-energy isomers of [C-benzene]$^+$, are significantly higher in energy than the 7C, 4C, and the cumulene.

\subsection{Energies of neutral  [C-coronene] complexes }
\label{sec:3.3Ccoronene}
It is well known that interstellar PAHs change their ionization through photoionization and electronic recombination much more frequently than C$^+$ accretion and 
even reactions with atomic H (Fig.\ \ref{fig:6A1-rates}). The same should be true for their carbon complexes, such that the stability of the various isomers of neutral 
[C-coronene] complexes must be considered in parallel to their cations. As seen in Fig. \ref {fig:4-figure4}, we considered bridging   each of the symmetry-unique CC bonds of coronene (N1-N4), insertion into 
CC bonds to give 7C rings (N5-N6), insertion into the CH bond (N9-N10), double CH insertion to give a 4C ring (N11) as well as pentagonal rings that we also considered   in the 
[C-coronene]$^+$ case (N12-N15). We note that neutral N14 turns into the spiro-cyclic isomer upon geometry optimization. Among the direct insertion products, 
a singlet 7C ring (N5) and the triplet CH-insertion (N10) products are particularly favourable, as is known for [C-benzene] (Fig.\ \ref{fig:1-figure1}). However, the most stable isomers are the 
singlet 4C ring (N11) that is without equivalent in [C-benzene] and the 5C ring with an exocyclic cumulene (N15).

\subsection{Energies of photodissociation products of [C-coronene] complexes}
\label{sec:3.4Ephotodiss} 

Figure \ref{fig:5-figure5} displays the energies of various dissociation products of   [C-coronene]$^+$. As expected, the products of H loss are significantly more stable than those of 
carbon loss, C$_x$H$_y$. Among the isomers of H loss, C$_{25}$H$_{11}^+$, the two most stable structures, with a heptagon ring (Ph1) or a pentagon ring with a ethynyl C2H group (Ph3), have 
practically equal energies. The dissociation energy for producing them from most stable C$_{25}$H$_{12}^+$, 3.1\,eV (Fig.\ \ref{fig:5-figure5}), is much less than H dissociation energy from 
 normal PAHs, $\sim$5\,eV. We may therefore expect that the limit in number of carbon atoms N$_{\rm C}$ for H-loss from [C-PAH]$^+$ cation complexes is  much higher than the value, 
$\sim$25-30, derived by Jochims et al.\ (1994, 1999) for  normal PAHs (Sect. \ref{sec:4.1summphotdiss}).

As expected, the photodissociation energies of C$_{25}$H$_{12}^+$ producing small carbonaceous fragments C$_x$H$_y$ (Ph4-Ph8) are generally much higher (Fig.\ \ref{fig:5-figure5}), including the loss of a 
neutral carbon atom (4.6\,eV). However, there is a channel with a dissociation energy for C$_2$H$_2$ loss from C6 of only  3.7\,eV to a structure with a 5C ring (Ph9). C$_2$H$_2$ loss from C14 requires even less energy  (2.1 eV, see Fig.\ \ref{fig:5-figure5}), while even   the most stable C15 loss of C$_2$H$_2$ requires only an additional 4.0 eV. Such channels might be 
considered among the processes for returning carbon to the ISM discussed in Sect. \ref{sec:4.2processingsmall}. We have been unable to identify a possible channel for such a fragmentation without an activation barrier that is  too high.

Similar photodissociation processes may occur for [C-PAH] neutral complexes, probably with somewhat higher photodissociation energies, as inferred from [C-benzene] 
complexes (Bettinger et al.\ 2000; Izadi et al.\ 2021). It is not expected that they significantly increase the rate of return of carbon to the ISM provided 
by  [C-PAH]$^+$ cation complexes (Sect. \ref{sec:4.2processingsmall}).

\subsection{Complexes of C and C$^+$ with dominant  interstellar PAHs}
\label{sec:3.5dominantPAHs}

It is known that coronene is probably too small to be stable in the diffuse ISM. However, it seems that the structure of its C-complexes may be representative of C-complexes of heavier stable pericondensed PAHs ('grandPAHs'),
 which should be dominant in interstellar conditions (e.g.\ Roelfsema et al.\ 1996; Tielens 2013; Andrews et al.\ 2015; Bouwman et al.\ 2019). Therefore, in the following section we   consider the energetics of the virtual fragmentation of these [C-coronene] complexes through H-loss and C$_2$H$_2$ loss in order to discuss the actual photofragmentation of heavier PAH complexes through similar processes in the ISM. We   assume that the energies of these processes are similar to coronene, since they should imply carbon rings at the PAH periphery with structures similar to coronene.

\section{Photodissociation of small [C-PAH] complexes in the diffuse ISM}
\label{sec:4smallCPAH}

\subsection{Summary  of  PAH photodissociation}
\label{sec:4.1summphotdiss}
There is a long history of the investigation of interstellar PAH photodissociation (e.g. Jochims et al.\ 1994, 1999; Allain et al.\ 1996a,b; Le Page et al.\ 2001, 2003; Tielens 2005, 2008, 2013; Montillaud et al.\ 2013, Andrews et al.\ 2016; Castellanos et al.\ 2018a,b; Stockett et al.\ 2023, 2024).   The photoionization experiments of Jochims et al.\ (1994, 1999) provide a   
basis for the threshold of the internal energy  needed for H loss as a function of the number of carbons, N$_{\rm C}$, in PAH cations. They yield that the direct single-photon photodissociation limit for H-loss is about N$_{\rm Clim}$\,=\,30-35 with the interstellar UV-photon energy lower than 13.6\,eV. However, this limit might be slightly lower, N$_{\rm Clim}$\,$\sim$\,25-30, when the contribution  of optical fluorescence, which accelerates the PAH cooling (e.g. Lacinbala et al.\ 2022; Bernard et al.\ 2023; Stockett et al.\ 2023, 2024),  is included in addition to infrared emission. We note, however, that Joblin et al.\ (2020) reported evidence for a hydrogen dissociation channel for large PAHs, with a very low branching ratio, through excited electronic states.
It is agreed that H loss is easier than C$_2$H$_2$ loss, so that, when possible, H loss protects PAHs from carbon loss. However, C$_2$H$_2$ loss has been observed in laser irradiation of small PAHs (e.g. Banhatti et al.\ 2022). Photodissociation of superhydrogenated PAHs (Sect. \ref{sec:4.2.4indirectC2H2loss}) may be easier.  
It had been proposed that Coulomb repulsion could favour photo fragmentation of PAH dications (Leach et al.\ 1986, 1989), but calculations of Malloci et al.\ (2007) have shown that this is not generally the case. 
Photodissociation of heavier PAHs may be important through simultaneous  two-photon absorption in regions with strong UV, such as PDRs (Montillaud et al.\ 1993), but not in the diffuse ISM (Fig.\ \ref{fig:6A1-rates}). 

Together with superhydrogenation,  aliphatic CH$_x$ sidechains attached to PAHs are more easily photodissociated than normal PAHs. This is true not only for H loss, but also for C$_2$H$_2$ fragmentation (e.g. Marciniak et al.\ 2021). 

As discussed in the references above, RRKM modelling of unimolecular reactions (e.g. Tielens 2005; Lange et al.\ 2021; Stockett et al.\ 2024) accounts for this photodissociation, assuming a complete thermalization of the injected photon energy. The Arrhenius form is a good approximation of RRKM results showing that the unimolecular dissociation rate roughly varies as exp(-E$_{\rm A}$/kT$_{\rm v}$) where E$_{\rm A}$ is  the activation energy and T$_{\rm v}$ is the microcanonical vibrational temperature. As recalled in Appendix \ref{app:CPhComplexes},
for PAHs and T$_{\rm v}$\,$<$\,1000\,K, T$_{\rm v}$ is related to N$_{\rm C}$ and the UV photon energy E$_{\rm UV}$ through the approximate relation (Eq. \ref{eq:C2Tv2}, inferred from Lange et al.\ 2021, updating Tielens 2005) 
\begin{equation}
        \label{eq:1Tv}
        {\rm T_{\rm v} \sim 2000 \times [E_{UV}(eV)/N_C]^{0.45} ~K}.
\end{equation}
Typical values of  E$_{\rm A}$ for pericondensed PAHs, which are thought to be most abundant among interstellar PAHs -- $\sim$5\,eV for H loss, higher than 6\,eV for C$_2$H$_2$ fragmentation (West et al.\ 2019) --  explain well the photodissociation behaviour discussed above. In particular, the absence of direct C$_2$H$_2$ photo-loss is confirmed for actual interstellar  PAHs  with N$_{\rm C}$\,$\ga$\,50 (West et al.\ 2019).

\subsection{Processing of small [C-PAH] complexes}
\label{sec:4.2processingsmall}

 As said in Sect. \ref{sec:3.5dominantPAHs}, we assume that we can infer the formation, evolution, and processing of small interstellar [C-PAH]$^+$ complexes from [C-coronene]$^+$ complexes. 

\subsubsection{H loss}
\label{sec:4.2.1Hloss}
For small [C-PAH]$^+$ complexes in the diffuse ISM,  the fastest chemical process (Fig. \ref{fig:6A1-rates}) is  H-loss (Fig.\ \ref{fig:5-figure5}) after the absorption of a $\sim$10\,eV UV photon.  
We assume that the 
 reaction energy for H-loss of [C-PAH]$^+$ complexes is comparable to that of the C6 isomer of [C-coronene]$^+$ (i.e.\ E$_{\rm A}$\,$\sim$\,3.1 eV; Fig.\ \ref{fig:5-figure5}). It is significantly smaller than for  normal PAHs, $\sim$4-5 eV (e.g. West et al.\ 2018; Castellanos et al.\ 2018a). As a rough guess, we assume that the ratio of the activation energies for H-loss of [C-PAH] complexes to those of normal PAHs is comparable to the ratio of their reaction energies (i.e.\ $\sim$3.1/(4-5).

We show  in Appendix \ref{app:C.1Scaling} (Eq. \ref{eq:C5NClim2}) that one may assume a rough dependence on E$_{\rm A}^{-2.2}$ for the N$_{\rm C}$ dehydrogenation limit (N$_{\rm ClimCcomp}$) of molecules with similar H-loss Arrhenius prefactors, such as PAHs and their carbon complexes. For [C-PAH]$^+$ complexes, this roughly yields a  factor of $\sim$[(4-5)/3.1]$^{2.2}$\,=\,1.75-2.85, i.e. N$_{\rm ClimCcomp}$ $\sim$45-80, compared to the limit $\sim$25-30\,C for direct H-loss from  normal PAHs . This represents a substantial fraction of stable interstellar PAHs, perhaps about one-half (see Appendix \ref{app:Dsize}). However, the actual value of N$_{\rm ClimCcomp}$ is very uncertain, in particular because of the unknown value of E$_{\rm A}$ for [C-PAH]$^+$ complexes. 
 
The value of N$_{\rm ClimCcomp}$ could be significantly lower for direct photo H-loss of neutral [C-PAH] complexes, reflecting higher H-dissociation energies (Sect. \ref{sec:3.3Ccoronene}); however, the higher efficiency of H loss from cations is transmitted to neutrals by electronic recombination.  Since the first H loss of small [C-PAH]$^+$ complexes has a rate comparable to UV-photon absorption and is an order of magnitude faster than typical H recombination in the diffuse ISM (Fig.\  \ref{fig:6A1-rates}), the majority of these carbon complexes with small interstellar PAHs should be partially dehydrogenated.

 The ease of H loss from carbon complexes is confirmed by the results of Canosa et al.\ (1995) for the reaction of C$^+$ + anthracene (C$_{14}$H$_{10}$), where the accretion product is not C$_{15}$H$_{10}^+$, but C$_{15}$H$_{9}^+$ with H loss. It is possible that the same occurs for C$^+$ + coronene or other small pericondensed PAHs, with H-loss products similar to Ph1 or Ph3 (Fig.\ \ref{fig:5-figure5}). We note that the internal energy of the complex similar to C6 or C15 before H loss is $\sim$9\,eV (i.e. comparable to that of a typical UV photon), confirming the high probability of photo H-loss for such photons and low binding energy of $\sim$3.1\,eV.

\medskip
\subsubsection{Isomerization and carbon loss}
\label{sec:4.2.2Closs}
Because of their relative instability, it is expected that carbon loss from carbon-PAH complexes, in the form of 
 C$_2$H$_2$, is significantly easier than from normal PAHs.
Several processes could lead to such a result because of the complexity of the photo-chemistry of these large molecules. Estimating the rate of carbon loss from all these processes is difficult.  
 C$_2$H$_2$ photo-loss from the various isomers of the [C-PAH]$^+$ and [C-PAH] complexes is discussed in Sect. \ref{sec:4.2.3directC2H2loss}, and from hydrogenated and dication complexes in Sect. \ref{sec:4.2.4indirectC2H2loss}.
        
Hydrogen  loss is generally much faster than carbon loss. However, 
if C$_2$H$_2$ loss is allowed, even with a very low branching ratio, br$_{\rm C2H2}$, it may be efficient in returning carbon to the interstellar gas. This is true when each H loss is rapidly compensated for by H re-accretion, so that after a large number of H loss-accretion cycles, the total C$_2$H$_2$ loss probability may be high before the next C$^+$ accretion by the PAH occurs. Therefore, direct photo-loss of C$_2$H$_2$ could contribute  to the decay of small [C-PAH]$^+$ complexes with a rate $\sim$10$^{-3}\times$ (br$_{\rm C2H2}/10^{-2}$)\,yr$^{-1}$, 
 which is faster than the rate of C$^+$ accretion $\sim$2$\times$10$^{-4}$\,yr$^{-1}$ (Fig.\  \ref{fig:6A1-rates}), 
if its average branching ratio br$_{\rm C2H2}$ proved to be significant (Appendix \ref{app:C.2directC2H2loss}).
        
        Because of the frequent charge exchange between neutral and cation complexes, carbon return from both species must be considered. The special case of dications is discussed in Sect. \ref{sec:4.2.4indirectC2H2loss}. Anions can probably be neglected because  photodetachment is relatively easy.
        
        A key question for the possibility of C$_2$H$_2$ photo-loss from carbon-PAH complexes is their isomer form. It may generally be assumed that they are in the form of the isomer with the lowest energy as the result of internal thermalization and low temperature. However, both for cation and neutral complexes, this lowest-energy isomer may be doubtful because of the uncertainty in the computed energies reported in Figs. \ref{fig:3-figure3} and \ref{fig:4-figure4}.
        
        In addition, it is not obvious to what degree isomer thermalization is achieved (at a temperature typically up to $\sim$ 1000\,K after a UV photon absorption or just after C$^+$ accretion, Eq. \ref{eq:1Tv}). It depends on the   activation energy for isomer transformations. Most of these activation energies are unknown for the various cation and neutral isomers of Figs. \ref{fig:3-figure3} and \ref{fig:4-figure4}, and the isomer transformation paths are uncertain.
\medskip

\subsubsection{C$_2$H$_2$ photo-loss from [C-PAH]$^+$ and [C-PAH] complexes}
        \label{sec:4.2.3directC2H2loss} 

It is likely that the first steps of C$^+$ accretion onto PAHs are similar to benzene (Sect. \ref{sec:2.1benzene}), leading to the initial formation of a 7C ring similar to compound C6 for  [C-coronene]$^+$ (Fig.\ \ref{fig:3-figure3}).  
 As quoted in Sect. \ref{sec:4.2.1Hloss}, it is probable that compounds similar to C6 immediately lose an H atom, yielding Ph1 or Ph3. However, these two compounds are fairly stable against carbon loss. Therefore, one must  discuss carbon loss after they return to C6 by H accretion.
As seen in Fig.\                                                                                                                                                                                                                                                                                                                                                                                       \ref{fig:5-figure5}, the dissociation energy for C$_2$H$_2$ extrusion from C6, 3.7 eV, is fairly low compared to more than 6\,eV for coronene (West et al.\ 2019), and only slightly higher than for H-loss, 3.1 eV. It seems that this result may be extended to all PAH cations, especially if pericondensed. However, the difficulty in locating the intermediate state for C$_2$H$_2$ extrusion from C6 [C-coronene]$^+$ suggests that the mechanism is likely not a single step, but more complex. The activation energy E$_{\rm A}$ and the branching ratio br$_{\rm C2H2}$  therefore remain uncertain.

The 7C-ring isomer C6 is not the most stable for [C-coronene]$^+$. The most stable form of this cation complex seems to be C15 with a 5C ring and an exocyclic cumulene. The energy difference with all other forms, $\ga$\,8\,kcal/mol, is significant and possibly greater than the energy uncertainties (Sect. \ref{sec:3CCpPAH}). Similarly, the lowest energy isomer of the neutral form of [C-coronene] seems to be N15 with a similar structure. It is therefore likely that the normal forms of [C-PAH]$^+$ and [C-PAH] complexes are similar to C15 and N15, respectively, after internal thermalization following absorption of UV photons. As quoted in Sect. \ref{sec:3.4Ephotodiss}, the C$_2$H$_2$ extrusion energy from C15 is only 4.0\,eV. This is not much higher than for H loss (3.3\,eV). It is thus possible that C$_2$H$_2$ photo-extrusion may occur in parallel to H photodissociation with a low branching ratio, if the activation energy is not too high. Isomer C14 might be considered as a possible transition state to C$_2$H$_2$ extrusion.  However, such a scheme for C$_2$H$_2$ extrusion from C15 would need to be confirmed by experiment or by computation of all activation energies and transition states. 

C$_2$H$_2$ extrusion might also be considered from the isomers of [PAH-C]$^+$ complexes similar to C9 or C10 for coronene (Fig.\ \ref{fig:5-figure5}), which have a computed energy only $\sim$10\,kcal/mol higher than C15 and $\sim$3\,kcal/mol higher than C6. It is not impossible that such isomers directly form during the accretion process. They might also be the result of some subsequent (photo)processes, such as photo-ionization or electronic recombination. For instance, one might consider that such isomers might be formed by photo-ionization of the bridged form of neutral [C-PAH] similar to N-11 (Fig.\  \ref{fig:4-figure4}). As this is  one of the most stable forms of [C-PAH], this channel might be very efficient. However, it may be quite difficult to prove such a conjecture.

 These [PAH-C]$^+$ complexes, similar to C9 or C10, might also be considered as intermediate steps in C$_2$H$_2$ extrusion from other more stable isomers. In this respect, it is important to note that Marciniak et al.\ (2021) have measured high branching ratios
 for C$_2$H$_2$ photo-extrusion from several compounds coronene-CH$_{\rm x}^+$ and pyrene-CH$_{\rm x}^+$. This includes (i) coronene-CH$_{\rm 2}^+$ (i.e.\ hydrogenated C9 or C10 compounds, C$_{25}$H$_{13}^+$, and pyrene-CH$_{\rm 2}^+$, C$_{17}$H$_{11}^+$, where C$_2$H$_2$ photo-extrusion branching ratios reach 70\% and 62\%, respectively; see  next section) and (ii) C$_{17}$H$_{9}^+$ with a branching ratio equal to 32\%. Although Marciniak et al.\ (2021) give no information for the actual isomer form that they observe for C$_{17}$H$_{9}^+$, nor for the photodissociation of coronene carbon complex, C$_{25}$H$_{13}^+$ (i.e.\ one of the dehydrogenated forms, Ph1, Ph2, or Ph3 of C6, C9, for example), 
their study shows that  C$_{25}$H$_{13}^+$ can have a significant C$_2$H$_2$ photo-extrusion branching ratio.
Other C$_2$H$_2$ extrusion channels might be considered, including from neutral complexes.  
Considering hydrogenated and dication complexes opens additional possibilities, as discussed in the next section.

\medskip
\subsubsection{Carbon loss from hydrogenated complexes}
        \label{sec:4.2.4indirectC2H2loss}

 Photodissociation of (singly) hydrogenated [HC-PAH]$^+$ (or [HC-PAH]) complexes might also be considered because there is a small probability $\eta$ (at most a few percent) of H accretion by [C-PAH]$^+$  before its H-loss photodissociation (see Fig.\ \ref{fig:6A1-rates}). For coronene, this gives C$_{25}$H$_{13}^+$. It is possible that the most stable and probable form of this compound is the isomer with an exocyclic CH$_2$ group. As quoted above, it is known that its photodissociation leads to carbon loss through C$_2$H$_2$ (or C$_2$H) extrusion and formation of a 5C ring with a probability close to 80\% (Marciniak et al.\ 2021). 
We may assume that photodissociation yields a similar result for  similar larger pericondensed H[C-PAH]$^+$ complexes up at least to N$_{\rm C}$ $\sim$ 50 (Appendix \ref{app:C.3Hydrogenation}).  While H photodissociation of [C-PAH]$^+$ strongly dominates over H accretion, the resulting dehydrogenated complex (Ph1 or Ph3, Fig. \ref{fig:5-figure5}) finally returns to [C-PAH]$^+$ by accreting a new H. 
However, it is difficult to infer the overall rate of C$_2$H$_2$ loss from this process because it depends on the complex cycle of H losses and re-accretions until the next formations of PAH-CH$_2^+$. Nevertheless, it might substantially contribute to the return of C$_2$H$_2$ to the interstellar gas from relatively small PAHs, if similar isomers may form in the processing of carbon PAH complexes.

Superhydrogenation of PAHs before the C$^+$ accretion (Appendices \ref{app:BsuperH} and \ref{app:C.3Hydrogenation})   
might also increase the instability of the carbon complex and favour C$_2$H$_2$ loss if superhydrogenation survived C$^+$ accretion, but to date precise data are lacking. 

Double ionization of the PAHs might also be considered.   
The first and second ionization potentials (IP) are lowered for heavier PAHs. Zhen et al.\ (2015, 2016) and Wenzel et al.\ (2020; see also Bern\'e et al.\ 2022) have shown that as the size increases, ionization dominates for large PAH cations after UV absorption above the second IP\,$\la$\,8\,eV. 
Double ionization was proposed as a fragmentation path by Coulomb explosions for normal PAHs by Leach et al.\ (1986, 1989) and through dissociative recombination by Millar et al.\ (1995), but its effects appear uncertain for normal PAHs (Malloci et al.\ 2007). 
Dissociative recombination might be more efficient for fragmenting dication carbon-PAH complexes, favouring C$_2$H$_2$ extrusion from various isomers. However, this should be confirmed by a detailed analysis of the various possible channels, whose activation energies are unknown. 

On the whole, despite the uncertainty of each of these possible processes leading to C$_2$H$_2$ extrusion from [C-PAH]$^+$ or [C-PAH] complexes, their number and diversity seem to guarantee
that most C$^+$ accretions onto small normal interstellar PAHs, with N$_{\rm C}$ below a limit N$_{\rm Clim-C2H2}\ga$\,50, should quickly return one C$_2$H$_2$ (or C$_2$H) molecule to the gas ISM before the next C$^+$ accretion. Depending on the value of N$_{\rm Clim-C2H2}$, the fraction of 
accreted carbon directly releasing a C$_2$H$_2$ molecule might reach 50\% or more, following the PAH size distribution. 

\bigskip
\section{Extension to large PAHs}
\label{sec:5largePAHs}

For any of the processes considered above, it seems likely that, beyond a certain number of carbon atoms, N$_{\rm Clim-C2H2}$, the energy of a single interstellar  UV photon, $<$\,13.6\,eV, is too low to allow the direct photo carbon-loss of the carbon complexes [C-PAH] formed by the accretion of a single C$^+$ ion onto pericondensed PAHs because their internal vibration temperature is too low. 
The value of  N$_{\rm Clim-C2H2}$ is uncertain, but is probably $\ga$40-50.
For N$_{\rm C}$\,$>$\,N$_{\rm Clim-C2H2}$, the existence of PAH carbon complexes with several additional carbon atoms  therefore seems possible. 
Even the continuous growth of carbon complexes  by successive C$^+$ accretions cannot be excluded. We note, however, that C$^+$ cannot accrete onto cations and that its rate of accretion onto large PAHs (and PAH clusters) could be drastically reduced if they were mostly positively charged, as are interstellar dust grains in the diffuse ISM.

Of course, there is no doubt that globally carbon is returned to the interstellar gas at the same rate as C$^+$ accretion in order to preserve the high abundance of C$^+$ in the gas. Therefore, some processes   certainly also exist for returning the accreted carbon from unstable complexes formed by the largest interstellar PAHs with N$_{\rm C}$\,$>$\,N$_{\rm Clim-C2H2}$.  
Several alternatives seem possible for the way  complexes with two or more additional carbon atoms could  return carbon to the gas.
Many of them are common to small PAHs (Sect. \ref{sec:4smallCPAH} and Appendix \ref{app:CPhComplexes}) and they could also ensure a prompt return of carbon  to the interstellar gas in the form of small molecules, mostly ${\rm C_{n}H_{m}}$ (mainly n\,=\,2, especially C$_2$H$_2$). Although we have not been able to precisely identify any specified process, we   briefly discuss them generically in the context of large PAHs.

\begin{figure}[htbp]
        \begin{center}
                \includegraphics[scale=0.7, angle=0]{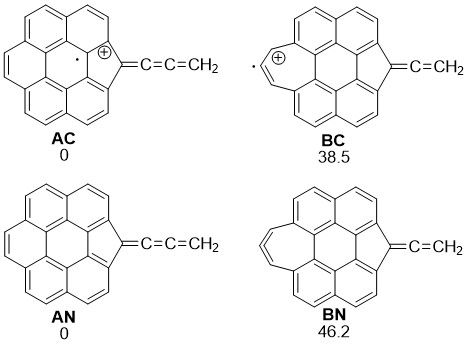}
                \caption{Relative energies [kcal/mol] of products of reactions of C15 and N15 (Figs.\ \ref{fig:3-figure3} \& \ref{fig:4-figure4}) with additional carbon, as computed at B3LYP/6-311+G** (see Sect. \ref{sec:3.1method}).
			}
                \label{fig:6-figure6}
        \end{center}
\end{figure}

\smallskip
-- Open photodissociation channels with low branching ratios

We have seen that a typical PAH absorbs more than 10$^4$ UV photons between two successive C$^+$ accretions. Therefore, any open channel for carbon release, even with an extremely low branching ratio, should be efficient for the prompt return of carbon to the gas. The inclusion of complexes with two or three
additional C atoms could open new extrusion channels in addition to the processes discussed for single carbon accretion. 
Such possible processes should be multiplied by the growth process of the PAH carbon skeleton, that inevitably includes unstable structures, such as loosely bound peripherical 4C, 5C, or 6C rings or singly bonded adducts. The direct photo-excitation of dissociative molecular levels, without thermalization, should also be considered. 

\smallskip
 -- Hydrogenation and multi-step photodissociation of [C-PAH] complexes

Larger PAHs multiply the number of isomers of [C-PAH] complexes and of sites for additional H atom bonding. 
This increases the chance of decreasing carbon-loss activation energies by (super)hydrogenation. 

\smallskip

-- Oxygenation

All previous processes favour a prompt return of carbon to the interstellar gas, mostly in the form of small molecules, ${\rm C_{n}H_{m}}$, after the accretion of a few C$^+$ ions. Nevertheless, if the carbon growth of the complexes continued,  one could consider the sputtering effect of oxygenation, although the accretion rate of O atoms is slower than C$^+$ ions by an order of  magnitude (Fig.\ \ref{fig:6A1-rates}). The most likely product of O accretion is the release of a CO molecule (Sect. \ref{sec:7CO}).
 However, larger photo fragments ${\rm C_{n}H_{m}}$O are not excluded.   

\smallskip
-- Possibility of carbon-chain adducts

        Chain formation as adducts of large interstellar PAHs appears possible  (see e.g.\ Zanolli et al.\ 2023; see also chain and ring formation in laser processing of PAHs, Panchagnula et al.\ 2024), since they are among the most stable [nC-PAH] complexes and they could easily form without high barrier from C-roaming on the PAH surface eased by photon absorption. Their length could be mostly limited either by ring formation from the attachment of their end to the PAH or by fast destruction.   
        In order to support the possibility that such chain adducts might be among the most stable forms of the addition of a few C atoms onto PAHs, we have computed the energies of cation and neutral compounds AC and AN (Fig.\ \ref{fig:6-figure6}) formed by attachment of a 3C cumulene chain to coronene through a 5C ring, and we have compared them with the energies of BC and BN, respectively, formed by the addition of a C atom to C15 and N15, respectively, forming a 7C ring (Fig.\ \ref{fig:6-figure6}). The differences in energies are substantial in favour of the compounds with a 3C cumulene sidechain: E(BC)-E(AC) = 39\,kcal/mol 
                and E(BN)-E(AN) = 46\,kcal/mol. Although similar comparisons should be performed with all possible structures alternative to sidechains, these results appear promising 
                for searching a confirmation of the importance of chain growth on large interstellar PAHs.
         
        While these chain adducts might be stabilized by fast recurrent fluorescence, various processes might be able to induce their detachment from the PAH, such as double ionization, oxygenation, or more energetic processes (see below).
        All these processes are too uncertain to allow any solid conclusion. However, the possibility of such chain adducts and of the fact that they feed the ISM with free long chains deserves a dedicated study, which  is beyond the scope of the present paper. 

It should be noted,  if one were tempted to consider higher energy processes to dissociate carbon complexes of large PAHs, that none seems fast enough to compensate for C$^+$ accretion.  Two-photon absorption is extremely rare in the diffuse ISM, about every 3x10$^7$\,yr for a given PAH (Fig.\ \ref{fig:6A1-rates}), which is about 10$^4$ times slower than C$^+$ accretion, while it is important in PDRs (e.g. Montillaud et al.\ 2013). The rate of absorption of extreme-UV photons, mostly in the range 400-900\,\AA\ (e.g.\ Fig.\ 1.9 of Tielens 2005), with a cross-section of $\sim$10$^{-17}$cm$^2$, is about 10$^2$ times slower than C$^+$ accretion. The destruction rate of PAHs, mostly by cosmic rays or in shocks, is $\sim$10$^4$-10$^5$ times slower than C$^+$ accretion (Micelotta et al.\ 2010a,b, 2011; see also Hrodmarsson et al.\ 2023, 2022).

To summarize, because of the complexity of interstellar PAHs and their physics, it is clear that  the discussion of the way they return carbon accreted as C$^+$ to the interstellar gas cannot be entirely conclusive. It is likely, however, that the main product they inject into the gas should be C$_2$H$_2$ because it is the most stable small carbonaceous fragment that they can produce.  This seems clear for the relatively small interstellar PAHs, with N$_{\rm C}$\,$\la$\,50-60, which may include more than half of all PAHs,  for example as supported by the James Webb Space Telescope (JWST) results of Chown et al.\ (2024).
For such small PAHs we can identify a number of possible channels for C$_2$H$_2$ release after repeated single-photon absorptions (Sect. \ref{sec:4.2.3directC2H2loss}). ${\rm C_{n}H_{m}}$ loss (mostly with n=2 and m=2) also seems the most probable output of C$^+$ accretion by larger PAHs with 70\,$\la$\,N$_{\rm C}$\,$\la$\,150 (Sect. \ref{sec:5largePAHs}), although it is more difficult to disentangle the actual processes. 

We can therefore conjecture that the majority of C$^+$ accretions likely lead to the injection of small carbonaceous molecules ${\rm C_{n}H_{m}}$ into the gas (together with a small fraction of CO molecules, Sect. \ref{sec:7CO}), and that C$_2$H$_2$ is vastly dominant among these carbonaceous molecules, with a fraction probably greater  than  50\%, so that, from Eq. \ref{eq:D5RCp}, the rate of injection of C$_2$H$_2$ molecules  (equal to one-half the rate of their carbon atoms R$_{\rm C}$) should be 
\begin{equation}
        \label{eq:2C2H2}
{\rm R_{C2H2}>\sim 2\times10^{-17}~cm^{-3}s^{-1}}.
\end{equation}
Such  photo-processing of [carbon-PAH] complexes could be complemented by  the processing of normal putative aliphatic PAH derivatives, which was suggested as an abundant source of small hydrocarbons by Marciniak et al.\ (2021).  However, it is possible that C$^+$ accretion is the most important source of aliphatic PAH derivatives.
As seen in Fig.\ \ref{fig:5-figure5}, an additional effect should be the efficient formation of stable PAHs that contain a peripheral pentagonal ring (see Sect. \ref{sec:8.2pentagons}).

\section{CO  sputtering through atomic O reactions with PAH cations}
\label{sec:7CO}
Atomic oxygen easily reacts with PAH cations (Betts et al.\ 2006; Snow and Bierbaum 2008). The possible effects in the diffuse ISM are reviewed in Appendix A.3 of Omont \& Bettinger (2021). They may be summarized as follows.  Pure accretion is expected onto interstellar PAHs with a rate of at least 1.3$\times$10$^{-10}$\,cm$^3$s$^{-1}$. 
This yields  about 3$\times$10$^{-5}$\,yr$^{-1}$ for the  rate of oxygen accretion onto a PAH in the standard diffuse ISM (Fig.\ \ref{fig:6A1-rates}). 
It is probable that an epoxy binding on two adjacent C-C carbons 
is initially formed. However, such a weak epoxy binding  cannot survive the 
next UV photon absorption unless the O atom forms a more stable binding. 
This is favoured by its high mobility on the PAH surface, so that a ketone appears as a possible outcome. For low values of  N$_{\rm C}$, the ketone may be easily photodissociated with loss of CO, whose dissociation energy and barrier may be lower than 2\,eV and 4\,eV, respectively, as found by Chen et al.\ (2018) for bisanthenquinone. The net effect of reactions with O for relatively small stable PAHs, with N$_{\rm C}$\,$<$\,50-60, should therefore be carbon sputtering of PAHs with injection of CO into the gas. 

It is likely that this result may be extended to all interstellar PAHs.
However, the effect of PAH oxidation on the PAH size distribution and on the global rate of carbon return to the gas is probably not important because the rate of accretion is an order of magnitude faster for C$^+$ than for O (Fig.\ \ref{fig:6A1-rates} and Eqs. \ref{eq:D5RCp}-\ref{eq:D6RO}). The carbon loss through CO loss is easily compensated for by the next C$^+$ accretion. Nevertheless, CO should be returned to the gas with a rate close to the O accretion rate (Eq. \ref{eq:D6RO}): 

\begin{equation}
        \label{eq:3RCO}
{\rm R_{CO} ~\sim 10^{-17} ~~ cm^{-3}s^{-1}}.
\end{equation} 
This should contribute  a minor part ($\sim$5\%) to the carbon return to the interstellar gas from C$^+$ accretion onto PAHs.
It is also shown, in Appendix \ref{app:E.2CO}, that this process is only a marginal source of formation of CO in the diffuse ISM.

\section{Astrophysical consequences}
\label{sec:8astrophysics}

\subsection{Efficient carbon return to the interstellar gas}
\label{sec:8.1Creturn} 

Our main conclusion supports the conjecture that most of the carbon from C$^+$ accretion is probably rapidly returned from interstellar PAHs to the gas after less than a few thousand years. 
This is consistent with the stability of the observed balance between C$^+$ and PAHs in the diffuse ISM.
Therefore, it is not obvious that frequent C$^+$ accretion has an important impact on the properties of interstellar PAHs, their abundances and size distribution, in addition to the formation of a significant fraction of anomalous PAHs with a pentagonal ring  (Sect. \ref{sec:8.2pentagons}), and some secondary effects such as  efficient compensation for O sputtering (Sect. \ref{sec:7CO}) and/or  the existence of a minor fraction of PAHs in the form of [C-PAH] or [2C-PAH] complexes, with a possible significant fraction of aliphatic or cumulene PAH derivatives.

It would be important to extend such a study of C$^+$ accretion to interstellar carbonaceous nano-particles heavier than PAHs generally called very small grains (VSGs, Desert et al.\ 1990) and possibly dominated by PAH clusters (e.g.\ Rapacioli et al.\ 2006; Joblin et al.\ 2017).  Because of the larger particle size and the lower vibration temperature (Eq. \ref{eq:1Tv}), carbon loss should not be possible with a single UV photon. Even two-photon desorption might be inefficient because several hundred carbon atoms might accumulate before it takes place (Fig.\ \ref{fig:6A1-rates}). Therefore, one should consider different problematics, including cluster dissociation (Rapacioli et al.\ 2006), PAH destruction (Micellotta et al.\ 2010a,b, 2011; Zhen et al.\ 2014; West et al.\ 2014; Hrodmarsson et al.\ 2023, 2022) and carbon circulation in the ISM including dust grains (e.g. Jones et al.\ 2014). 
For PAHs in this context, it would be crucial to model their top-down evolution in detail (Zhen et al.\ 2014), including  formation from heavier particles, clusters, or grains; photosymmetrization into grandPAHs (Andrews et al.\ 2015; Bouwman et al.\ 2019); or photo downsizing, dehydrogenation (e.g.\ Omont \& Bettinger 2021), and eventual destruction (e.g.\ Hrodmarsson et al.\ 2023, 2022; Sundararajan et al.\ 2024; Panchagnula et al.\ 2024). 

\subsection{Increasing the structure variety of interstellar PAHs}
\label{sec:8.2pentagons}

 It is noted in Sects. \ref{sec:4.2.2Closs} and \ref{sec:8.1Creturn}
   that an important side effect of C$^+$ accretion on small interstellar PAHs and the subsequent C$_2$H$_2$ loss  is the efficient formation of  PAHs that contain a peripheral pentagonal ring. Such structures also form in the photodissociation of aliphatic PAH derivatives, and it is known that they can be as stable as, or even more stable than, the bare PAH cations (Marciniak et al.\ 2021). If C$_2$H$_2$ is ejected long before the arrival of the next C$^+$ ion, as such a structure should be stable, the fraction of PAHs with a pentagonal ring should be substantial.
 
 Other various relatively stable shapes of [C-PAH] complexes (e.g. with a heptagonal ring) 
may further increase the variety of interstellar PAHs. 
 Such structures, with C-rings other than hexagonal or adducts, might be also be  important for large PAHs if the carbon release is not achieved after the accretion of the first C$^+$, but from 2C or multi-C complexes.
We note that all such structures are much less symmetrical than canonical grandPAHs  (Andrews et al.\ 2015; Bouwman et al.\ 2019).   

\subsection{Carbon chemistry in the diffuse ISM and PDRs}
\label{sec:8.3chemistry}

As stated above, a key output of C$^+$ accretion onto PAHs is the frequent injection of small carbonaceous molecules, mainly C$_2$H$_2$, into the diffuse interstellar gas at a rate comparable to C$^+$ accretion (i.e.\ a few 10$^{-17}$\,cm$^{-3}$s$^{-1}$;  Eq. \ref{eq:2C2H2}). Together with the subsequent accretion of a second C$^+$ ion, one may consider this as C$_2$H$_2$ formation from PAH catalysis. It might be a key input for the chemistry of small carbon molecules, as anticipated especially from the observation of large abundances of small carbonaceous molecules, C$_{\rm x}$H$_{\rm y}$, in PDR boundary regions such as the Horsehead Nebula (Pety et al.\ 2005; Guzm{\'a}n et al.\ 2015).

As discussed in Appendix \ref{app:E.1abundances},
 from the rate of formation of C$_2$H$_2$ given by Eq. \ref{eq:2C2H2} and a typical value of the visible extinction A$_{\rm v}$\,=\,0.7, 
one may infer an abundance of C$_2$H in the diffuse ISM as (Eqs.\ \ref{eq:E2chiC2HnT}-\ref{eq:E3chiC2HH2})
\begin{equation}
        \label{eq:4chiC2H2}
{\rm n_{C2H}/n_{HT} \sim 3\times10^{-9}},
\end{equation} 
\begin{equation}
        \label{eq:5chiC2H}
{\rm n_{C2H}/n_{H2} \sim 7\times10^{-9}\times f_{H2}^{-1}},
\end{equation}
where  f$_{\rm H2}$ is the fraction of molecular hydrogen. This value -- ${\rm  n_{C2H}/n_{H2} \sim 2x10^{-8}}$ for  f$_{\rm H2}$\,$\sim$\,0.4 -- is  comparable to the universal value of the abundance of C$_2$H observed  in the dense diffuse ISM (n$_{\rm HT}$\,$\sim$\,100-1000 cm$^{-3}$) by Gerin et al.\ (2010), ${\rm n_{C2H}/n_{H2} \sim 3.2\pm 1.1\times 10^{-8}}$.

As detailed by Godard et al.\ (2009, 2014), it is possible to account for such large abundances of C$_2$H and related small carbonaceous molecules in the diffuse ISM by turbulent dissipation chemistry. However, we show that the contribution of PAHs in the formation of these molecules through C$_2$H$_2$ injection should be substantial. 

As suggested by  Pety et al.\ (2005) and Guzm{\'a}n et al.\ (2015), the same process should also be efficient in PDR boundary regions such as the Horsehead Nebula. However, our detailed estimates should be adapted to physical conditions, that are different in PDRs compared to the diffuse ISM. 

Although each O accretion onto PAHs$^+$ probably results in the injection of one CO molecule into the gas, CO loss from oxygenated PAHs seems a less important astrophysical process. Its contribution to the carbon return from PAHs to the interstellar gas is a  factor $\ga$4  smaller than C$_2$H$_2$ loss (Eqs.\ \ref{eq:2C2H2} and \ref{eq:3RCO}). In addition, as shown in Appendix D.2, 
the rate of formation of CO resulting from O accretion onto PAHs$^+$  is too slow by an order of magnitude to account for the CO formation rate inferred from the observed abundances of CO and HCO$^+$ in the diffuse ISM (e.g. Gerin \& Liszt 2021).

\section{Summary and conclusion}
\label{sec:9conclusion}

C$^+$ accretion onto PAHs is a key process for the carbon chemistry of the UV-irradiated ISM. The clarification of the resulting circulation of interstellar carbon between the diffuse gas and PAHs  remains difficult because of the complexity of PAHs and their chemical processes. There is no doubt that globally carbon is returned to the interstellar gas at the same rate as C$^+$ accretion onto PAHs in order to preserve the observed high abundance of  C$^+$ in the gas. 

It seems confirmed that the relative instability of carbon complexes formed from C$^+$ accretion onto interstellar PAHs, favours a prompt return of carbon to the interstellar gas. 
We have identified several channels through which about half of the accreted carbon is likely returned to the gas directly after its accretion on small PAHs before the accretion of the next C$^+$ ion.
        The loose structure of these complexes, including ones carrying aliphatic or cumulene side chains, 
        might allow the opening of several C$_2$H$_2$-loss photodissociation channels, at least  for small PAHs. 
        Even those  channels with low branching ratio compared to H-loss  might eventually approach 100\% C$_2$H$_2$ return to the gas after each C$^+$ accretion onto small normal PAHs and multiple cycles of photo H-loss and re-accretion (Sect. \ref{sec:4.2processingsmall}). 

The case of the evolution of larger pericondensed [C-PAH] complexes is less clear because their direct single-photon photodissociation seems difficult. However, it is possible that various  photo-processes or their combination can achieve the fragmentation of [C$_n$-PAH] complexes, with the release of small  ${\rm C_{n}H_{m}}$ molecules (mostly n=2 and m=2). The possibilities of photodissociation channels with low branching ratios, helped by hydrogenation, metastable isomer intermediates, or double ionization, appear promising. It would be important to confirm the details of such processes in order to better understand the interstellar PAH size distribution and its origin. 

 The identification of the most efficient channels for C$_2$H$_2$ photodissociation loss would require complex calculations of all reaction barriers. Although this is beyond  the scope of this paper, the multiplicity of these possible channels makes it likely that C$_2$H$_2$ loss occurs after each C$^+$ accretion onto small normal PAHs. Moreover, the mere fact of the stable high abundance of C$^+$ in the diffuse ISM means that carbon  accretion onto PAHs is compensated for by some carbon loss process. Our discussion shows that because of its relative stability, C$_2$H$_2$ loss is very probably the most efficient carbon-loss process from [C-PAH] complexes. Altogether, even in the absence of calculations of reaction barriers, it is highly likely that the rate of injection of C$_2$H$_2$ into the diffuse interstellar gas from PAHs is a substantial part of the rate of accretion of C$^+$ onto PAHs.

It seems confirmed, therefore, that PAHs substantially contribute to  the (catalytic) formation of small carbonaceous molecules in the diffuse ISM and moderate PDRs, mostly through injection of C$_2$H$_2$ as a key input for the chemistry of small carbon molecules.   

An additional effect of C$^+$ accretion on interstellar PAHs and the subsequent C$_2$H$_2$ extrusion  should be
 the efficient formation of very stable, but temporarily existing, anomalous PAHs that contain a peripheral pentagonal ring. As they should survive until the next C$^+$ accretion, the fraction of PAHs with a pentagonal ring should be  significant.              
        Together with the various [carbon-PAH] complexes, this should substantially increase the variety of interstellar PAHs and reduce their overall symmetry.

Although practically every O accretion onto PAHs$^+$ probably results in the injection of one CO molecule into the gas, CO loss from oxygenated PAHs seems a less important astrophysical process. It is only a minor contribution to carbon return from PAHs to the gas and to the formation of CO in diffuse clouds.

The evolution of complexes of carbon with catacondensed PAHs, especially polyacenes, and their possible resulting growth  should also be investigated, despite its complexity. One might also explore  whether similar processes may operate for PAHs in other astrophysical contexts, such as translucent clouds, PDRs, planetary nebulae, or preplanetary systems, and whether carbon chains may efficiently form and grow as PAH adducts (Sect. \ref{sec:5largePAHs}).

\bigskip
\begin{acknowledgements} 
We are indebted to Pierre Cox for his careful reading of the manuscript and his 
suggestions for improving it. 
We specially thank Christine Joblin for important suggestions. We thank Maryvonne G\'erin, Helgi Hrodmarsson, Ugo Jacovella and Evelyne Roueff for  helpful discussions.
The authors acknowledge support by the state of Baden-Württemberg through bwHPC
and the German Research Foundation (DFG) through grant no INST 40/575-1 FUGG (JUSTUS 2 cluster). 
\end{acknowledgements}

\bigskip

{\bf References}  
\smallskip

Allain, T., Leach, S., \& Sedlmayr, E.\ 1996a, \aap, 305, 602

Allain, T., Leach, S., \& Sedlmayr, E.\ 1996b, \aap, 305, 616

Allamandola, L.~J., Tielens, A.~G.~G.~M., \& Barker, J.~R.\ 1985, \apjl, 290, L25 

Andrews, H., Boersma, C., Werner, M.~W.\ et al.\ 2015, \apj, 807, 99 

Andrews, H., Candian, A., \& Tielens, A.~G.~G.~M.\ 2016, \aap, 595, A23 

 Baboul, A.~G., Curtiss, L.~A., Redfern, P.~C., \& Raghavachari, K.\ 1999, J.\ Chem.\ Phys., 110 7650

Bakes, E.~L.~O. \& Tielens, A.~G.~G.~M.\ 1994, \apj, 427, 822

Bakes, E.~L.~O., Tielens, A.~G.~G.~M., Bauschlicher, C.~W., et al.\ 2001, \apj, 560, 261

Banhatti, S., Rap, D.~B., Simon, A.\ et al.\ 2022, Phys. Chem. Chem. Phys., 24, 27343 

Bauschlicher, C. W., Jr., \& Ricca, A., 2014, Theor. Chem. Acc., 133, 1479

Becke, A.~D., Lee, C., Yang, W., \& Parr, R.~G.\ 1993, J.\ Chem.\ Phys., 98, 5648

Bernard, J., Ji, M., Indrajith, S., Stockett, M.~H.\ et al.\ 2023, Phys. Chem. Chem. Phys., 25, 10726
        
Bern{\'e}, O., Foschino, S., Jalabert, F., et al.\ 2022, \aap, 667, A159

Bettinger, H.~F., Schleyer, P.~v.R., Schaefer, H.~F.\ et al.\ 2000, J.\ Chem.\ Phys., 113, 4250

Betts, N.~B., Stepanovic, M., Snow, T.~P., \& Bierbaum, V.~M. 2006, \apjl, 651, L129

Black, J.~H., van Dishoeck, E.~F., Willner, S.~P., et al.\ 1990, \apj, 358, 459

Bouwman, J., Castellanos, P., Bulak, M., et al.\ 2019, \aap, 621, A80

Canosa, A., Laub\'e, S., Rebrion, C., Pasquerault, et al.\ 1995, Chem.\ Phys.\ Lett., 245, 407 

Cardelli, J.~A., Meyer, D.~M., Jura, M., \& Savage, B.~D.\ 1996, \apj, 467, 334

Castellanos, P., Candian, A., Andrews, H., \& Tielens, A.~G.~G.~M.\ 2018a, \aap, 616, A166 

Castellanos, P., Candian, A., Zhen, J., Linnartz, H., \& Tielens, A.~G.~G.~M.\ 2018b, \aap, 616, A167

Chen, T., Zhen, J., Wang, Y., Linnartz, H., \& Tielens, A. G. G. M. 2018, Chem.\ Phys.\ Lett., 692, 298

Chown, R., Sidhu, A., Peeters, E., et al.\ 2024, \aap, 685, A75

Dalgarno, A. \& Black, J.~H.\ 1976, Reports on Progress in Physics, 39, 573

Desert, F.-X., Boulanger, F., \& Puget, J.~L.\ 1990, \aap, 237, 215

Draine, B.~T. \& Li, A.\ 2007, \apj, 657, 810

Frisch, M.~J., Trucks, G.~W., Schlegel, H.~B., et al. 2016, Gaussian~16 Revision C.01, gaussian Inc. Wallingford CT

Gerin, M., de Luca, M., Goicoechea, J.~R., et al.\ 2010, \aap, 521, L16

Gerin, M., Neufeld, D.~A., \& Goicoechea, J.~R.\ 2016, \araa, 54, 181

Gerin, M. \& Liszt, H.\ 2021, \aap, 648, A38

Godard, B., Falgarone, E., \& Pineau Des For{\^e}ts, G.\ 2009, \aap, 495, 847

Godard, B., Falgarone, E., \& Pineau des For{\^e}ts, G.\ 2014, \aap, 570, A27

Guzm{\'a}n, V.~V., Pety, J., Goicoechea, J.~R., et al.\ 2015, \apjl, 800, L33

Hahndorf, I., Lee, Y.~T, Kaiser, R.~I., et al.\  J.\ Chem.\ Phys., 116, 3248

Hamberg, M., Kashperka, I., Thomas, R.~D., Roueff, E., et al.\ 2014, J.\ Phys.\ Chem.\ A, 118, 6034

Hrodmarsson, H.~R., Bouwman, J., Tielens, A.~G.~G.~M., \& Linnartz, H.\ 2022, International Journal of Mass Spectrometry, 476, 116834

Hrodmarsson, H.~R., Bouwman, J., Tielens, A.~G.~G.~M., \& Linnartz, H.\ 2023, International Journal of Mass Spectrometry, 485, 116996

Izadi, M. E., Bal, K. M., Maghari, A., \& Neyts, E.C. 2021, Phys. Chem. Chem. Phys., 23, 4205

Jensen, P.A., Leccese, M.,  Simonsen, F.D.S., 2019, MNRAS, 486, 5492

Joblin, C., Dontot, L., Garcia, G.~A., et al.\ 2017, Journal of Physical Chemistry Letters, 8, 3697

Joblin, C., Wenzel, G., Rodriguez Castillo, S., et al.\ 2020, Journal of Physics Conference Series, 1412, 062002

Jochims, H. W., R{\"u}hl, E., Baumg{\"a}rtel, H., Tobita, S., \& Leach, S. 1994, ApJ, 420, 307

Jochims, H. W., Baumg{\"a}rtel, H., \& Leach, S. 1999, ApJ, 512, 500

Jones, A.~P., Ysard, N., K{\"o}hler, M., et al.\ 2014, Faraday Discussions, 168, 313

Kaiser, R.~I., Hahndorf, I., Huang, L.~C.~L. et al.\ 1999, J.\ Chem.\ Phys., 110, 6091

Knight, C., Peeters, E., Stock, D.~J., et al.\ 2021, \apj, 918, 8

Krasnokutski, S.~A. \& Huisken, F.\ 2014, J.\ Chem.\ Phys., 141, 214306 

Krasnokutski, S.~A., Huisken, F., J{\"a}ger, C., et al.\ 2017, \apj, 836, 
32

Kwok, S.\ 2021, \apss, 366, 67

Lacinbala, O., Calvo, F., Dubosq, C., et al.\ 2022, \jcp, 156, 144305

Lange, K., Dominik, C., \& Tielens, A.~G.~G.~M.\ 2021, \aap, 653, A21

Leach, S.\ 1986, Journal of Electron Spectroscopy and Related Phenomena, 41, 421

Leach, S., Eland, J.~H.~D.,\& Price, S.~D.\ 1989, J.\ Phys.\ Chem., 93, 22, 7575, 7583

Leger, A. \& Puget, J.~L.\ 1984, \aap, 500, 279

Le Page, V., Snow, T.~P., \& Bierbaum, V.~M.\ 2001, \apjs, 132, 233 

Le Page, V., Snow, T.~P., \& Bierbaum, V.~M.\ 2003, \apj, 584, 316

Lepp, S. \& Dalgarno, A.\ 1988, \apj, 324, 553

Lepp, S., Dalgarno, A., van Dishoeck, E.~F., et al.\ 1988, \apj, 329, 418

Liszt, H., Gerin, M., Beasley, A., et al.\ 2018, \apj, 856, 151

Malloci, G., Joblin, C., \& Mulas, G.\ 2007, \aap, 462, 627

Maragkoudakis, A., Peeters, E., \& Ricca, A.\ 2020, \mnras, 494, 642

McElroy, D., Walsh, C., Markwick, A.~J., et al.\ 2013, \aap, 550, A36

Marciniak, A., Joblin, C., Mulas, G., et al.\ 2021, \aap, 652, A42

Micelotta, E.~R., Jones, A.~P., \& Tielens, A.~G.~G.~M.\ 2010a, \aap, 510, A36

Micelotta, E.~R., Jones, A.~P., \& Tielens, A.~G.~G.~M.\ 2010b, \aap, 510, A37 

Micelotta, E.~R., Jones, A.~P., \& Tielens, A.~G.~G.~M.\ 2011, \aap, 526, A52

Millar, T.~J., Roueff, E., Charnley, S.~B., et al.\ 1995, International Journal of Mass Spectrometry and Ion Processes, 149, 389

Montgomery, J.~A.\ Jr., Frisch, M.~J., Ochterski, J.~W., \& Petersson, G.~A.\ 1999, J.\ Chem.\ Phys., 110, 2822

Montgomery, J.~A.\ Jr., Frisch, M.~J., Ochterski, J.~W., \& Petersson, G.~A.\ 2000, J.\ Chem.\ Phys., 112, 6532

Montillaud, J., Joblin, C., \& Toublanc, D.\ 2013, \aap, 552, A15 

Omont, A.\ 1986, \aap, 164, 159. 

Omont, A. \& Bettinger, H.~F.\ 2021, \aap, 650, A193

Panchagnula, S., Kamer, J., Candian, A., et al.\ 2024, Physical Chemistry Chemical Physics (Incorporating Faraday Transactions), 26, 18557

Pety, J., Teyssier, D., Foss{\'e}, D., et al.\ 2005, \aap, 435, 885

Puget, J.~L. \& Leger, A.\ 1989, \araa, 27, 161

Rapacioli, M., Calvo, F., Joblin, C., et al.\ 2006, \aap, 460, 519

Roelfsema, P.~R., Cox, P., Tielens, A.~G.~G.~M., et al.\ 1996, \aap, 315, L289

Rybarczyk, D.~R., Stanimirovi{\'c}, S., Gong, M., et al.\ 2022, \apj, 928, 79
        
Smith, R.~D., \& DeCorpo, J.~J.\ 1976, J.\ Phys.\ Chem., 80, 2904

Smith, R.~D., \& Futrell, J.~H.\ 1978, International Journal of Mass Spectrometry and ionic physics, 26, 111

Snow, T.~P., \& Bierbaum, V.~M. 2008, Annu.\ Rev.\ Anal.\ Chem., 229

Stockett, M.~H., Bull, J.~N., Cederquist, H., et al.\ 2023, Nature Communications, 14, 395

Stockett, M.~H., Bull, J.~N., Cederquist, H., et al.\ 2024, Nature Communications, 15, 8443

Sundararajan, P., Candian, A., Kamer, J., et al.\ 2024, Physical Chemistry Chemical Physics (Incorporating Faraday Transactions), 26, 19332

Thaddeus, P. 1995, in Tielens, A.~G.~G.~M., \& Snow, T.~P.\ 1995, The Diffuse Interstellar Bands, Astrophysics and Space Science Library, 202, 369 (Kluwer)

Tielens, A.~G.~G.~M.\ 2005, The Physics and Chemistry of the Interstellar 
Medium (Cambridge University Press, Cambridge, UK)

Tielens, A.~G.~G.~M.\ 2008, \araa, 46, 289 

Tielens, A.~G.~G.~M.\ 2013, Reviews of Modern Physics, 85, 1021

Watson, W.~D.\ 1978, \araa, 16, 585

Wenzel, G., Joblin, C., Giuliani, A., et al.\ 2020, \aap, 641, A98

West, B., Useli-Bacchitta, F., Sabbah, H., Blanchet, V.,  Bodi, A., Mayer, P.~M., Joblin, C.~J.\ 2014, Phys.\ Chem.\ A, 118, 7824

West, B., Rodriguez Castillo, S., Sit, A., Mohamad, S., Lowe, B., Joblin, C., Bodi, A., \& Mayer, P.~M.\ 2018, Phys.\ Chem.\ Chem.\ Phys., 20, 7195

West, B., Lesniak., L.\ \& Mayer, P.~M.\ 2019, J.\ Phys.\ Chem.\ A, 123, 3569

Zanolli, Z., Malc{\i}o{\u{g}}lu, O.~B., \& Charlier, J.-C.\ 2023, \aap, 675, L9

Zhen, J., Castellanos, P., Paardekooper, D.~M., et al.\ 2014, \apjl, 797, L30

Zhen, J., Castellanos, P., Paardekooper, D.~M., et al.\ 2015, \apjl, 804, L7

Zhen, J., Rodriguez Castillo, S., Joblin, C., et al.\ 2016, \apj, 822, 113

\appendix

\section{PAH and [C-PAH] superhydrogenation}
\label{app:BsuperH}

Superhydrogenation binds one or several additional H atoms to normal PAHs, generally to a carbon atom already bound to an H at the periphery. Considering coronene,  the computed value of H binding energy for superhydrogenation of neutral coronene (Jensen et al.\ 2019) is much smaller for the first H ($\sim$1.3 eV) than for the second H ($\sim$3.9 eV) of an external C-C bond. Similar values are expected for typical interstellar PAHs, mostly independently of their charge.   
It is agreed that accretion reactions of atomic H with PAH cations are practically  barrierless (Le Page et al.\ 2001; Andrews et al.\ 2016), to the difference of neutrals (Jensen et al.\ 2019). Accretion of one H atom onto PAH cations should therefore be easy in the diffuse atomic ISM. Its rate was estimated  as $\sim$1.4 10$^{-10}$ cm$^{3}$ s$^{-1}$ by Le Page et al.\ (2001) and Andrews et al.\ (2016), which yields a time constant of $\sim$5 yr for PAHs$^+$ accreting H, with n$_{\rm HT}$ = 50 cm$^{-3}$, which is \ $\sim$10\,yr (Fig.\ \ref{fig:6A1-rates}) for a PAH on average, assuming that it spends half the time as ionized. 
  
However, UV photo-absorption is much faster in the diffuse ISM, with  time constant $\sim$0.1 yr (Fig.\ \ref{fig:6A1-rates}). Because of the very small  binding energy of the first H, it is probable that odd-number interstellar superhydrogenation cannot survive, even for large PAHs. However, a tiny fraction of interstellar PAHs ($\sim$1\%) should be singly superhydrogenated. As calculated by Bauschlicher \& Ricca (2014) for superhydrogenated pyrene, they could theoretically lose C$_2$H$_2$ with a binding energy only $\sim$3.9\,eV. But their rate of C$_2$H$_2$ loss by photodissociation should be  
roughly more than 10$^{11}$ times slower than for the loss of the extra H atom.

As regards double superhydrogenation, assuming an activation energy E$_{\rm A}$$\sim$3.9\,eV, equal to the binding energy, one may roughly infer a stability limit N$_{\rm Clim-superH}$ for the PAH size in the diffuse ISM by scaling the limit N$_{\rm C}$~30-35 for direct aromatic H-loss in normal PAHs with E$_{\rm A}$\,$\sim$\,4.8\,eV, found by Jochims et al. (1994). Assuming that the vibrational temperature T$_{\rm v}$ should scale as E$_{\rm A}$, Eq. \ref{eq:C2Tv2} should roughly yield N$_{\rm Clim-superH}$\,$\sim$\,(4.8/3.9)$^{2.22}\times 30$ $\sim$\,50, or, rather, N$_{\rm Clim-superH}$\,$\sim$\,40, taking into account the effect of delayed fluorescence (see Sect. \ref{sec:4.1summphotdiss}). 

To summarize, odd-H superhydrogenation cannot survive photodissociation for all interstellar PAHs. It should be the same for even-H superhydrogenation for small PAHs. On the other hand, for large PAHs, double-H superhydrogenation on external C-C bonds could be significant, albeit uncertain. There is no reason that similar rules do not apply to carbon-PAH complexes, although aliphatic complexes deserve a special discussion (Sect. \ref{sec:4.2.4indirectC2H2loss}).

\section{Photodissociation of [C-PAH] complexes}
\label{app:CPhComplexes}

Just after its creation by C$^+$ accretion (or eventually after rehydrogenation, Sect. \ref{sec:4.2.3directC2H2loss}),  the complex [C-PAH]$^+$ may be transformed by one of the processes that are displayed in Fig.\ \ref{fig:6A1-rates} together with the orders of magnitude of their rates in the diffuse ISM. Most of the latter are thought to be similar for normal PAHs and their carbon complexes. The dominant process is H-loss  photodissociation by far. A number of other processes are significant, when the PAH size allows it, 
including electronic recombination, second ionization,  H accretion (Fig.\ \ref{fig:6A1-rates}), and  C$_2$H$_2$ photo-loss (see below). We denote their rates (s$^{-1}$) as $\gamma_{\rm er}$, $\gamma_{\rm i2}$,  $\gamma_{\rm aH}$  
(Fig.\ 1), and $\gamma_{\rm dC2H2}$, respectively. Similarly,  $\gamma_{\rm dH}$  
is the rate of  photo H-loss. It should be close to the rate of UV-photon absorption, $\gamma_{\rm UV}\sim10$\,yr$^{-1}$ (Fig.\ 1), for N$_{\rm C}$\,$<$\,N$_{\rm Clim}$ for normal PAHs, and practically equal to $\gamma_{\rm UV}$ for superhydrogenated small PAHs.

\subsection{Scaling H loss from PAHs to carbon-PAH complexes}
\label{app:C.1Scaling}

After the absorption of an interstellar UV photon of energy E$_{\rm UV}$ $\sim$ 6--13.6\,eV, one may assume a complete thermalization of the injected photon energy among the vibration modes. For PAHs and T$_{\rm v}$\,$<$\,1000\,K, T$_{\rm v}$ is related to N$_{\rm C}$ through the approximate relation (Lange et al.\ 2021, Eq.\ 4; updating Tielens 2005) 
\begin{equation}
        \label{eq:C1Tv1}
 {\rm T_{\rm v} \sim 3750 \times [E_{UV}(eV)/(3N-6)]^{0.45} ~K},
\end{equation}
where N\,=\,N$_{\rm C}$+N$_{\rm H}$ is the total number of atoms in the PAH.  Assuming N\,$>>$\,1 and N$_{\rm C}$/N$_{\rm H}$\,$\sim$\,3 yields
\begin{equation}
        \label{eq:C2Tv2}
 {\rm T_{\rm v} \sim 2000 \times [E_{UV}(eV)/N_C]^{0.45} ~K};
\end{equation}

RRKM modelling of unimolecular reactions accounts for PAH photodissociation with such a temperature (Tielens 2005; Lange et al.\ 2021). In the Arrhenius form, which is a reasonably good approximation of RRKM results, the unimolecular dissociation rate writes 
\begin{equation}
        \label{eq:C3gammaphi}
{\rm \Gamma_\Phi \sim A\,exp(-E_{\rm A}/kT_{\rm v})},
\end{equation}
where E$_{\rm A}$ is  the activation energy, the prefactor A is proportional to the variation of entropy, and k is the Boltzmann constant.  

After the absorption of a UV photon, photodissociation occurs when it is faster than the cooling of the PAH, i.e.\ when   $\Gamma_\Phi$\,$>$\,$\Gamma_{\rm cool}$, the total PAH cooling rate (i.e.\ the sum of the infrared and the recurrent-fluorescence cooling rates, e.g. Lacinbala et al.\ 2022; Bernard et al.\ 2023; Stockett et al.\ 2023, 2024). Therefore, the lower limit of the number of carbons of PAHs that escape 
photodissociation,  N$_{\rm Clim}$, is approximately given by $\Gamma_\Phi$\,=\,$\Gamma_{\rm cool}$. Eqs.\ \ref{eq:C3gammaphi} and \ref{eq:C2Tv2} then yield 
\begin{equation}
        \label{eq:C4NClim1}
{\rm (E_{UV}/N_{Clim})^{0.45} \sim E_A/[2000k\times ln(A/\Gamma_{cool})]}
\end{equation}
or 
\begin{equation}
        \label{eq:C5NClim2}
{\rm N_{Clim} \sim C\times E_A^{-2.2}}
\end{equation}
where C is a constant equal to\\ 
${\rm E_{UV}(eV)\times  [2000k\times ln(A/\Gamma_{cool})]^{2.2}}$. 

\subsection{ Direct C$_2$H$_2$ loss from [C-PAH]$^+$ complexes}
\label{app:C.2directC2H2loss}

As discussed in Sect. \ref{sec:4.2.3directC2H2loss}, direct photo extrusion of C$_2$H$_2$ from some isomers of [C-PAH]$^+$ (Fig.\ \ref{fig:3-figure3}) might be competitive, depending on the activation energy.   
As quoted in Fig.\ \ref{fig:5-figure5}, for [C-coronene]$^+$, the dissociation energy for the easiest carbon loss (C$_2$H$_2$), from the most stable isomers, C15 and C6, 4.0\,eV and 3.7 eV, respectively, is fairly low compared to more than 6\,eV for coronene (West et al.\ 2019), and it is only slightly higher than for H-loss, 3.1-3.3 eV. It seems that this result may be extended to all PAH cations, especially pericondensed.

 However, the  activation energies E$_{\rm A}$ remain uncertain. They might be much higher than 4\,eV, meaning that these channels are negligible.
 Nevertheless, if E$_{\rm A}$ was not too high, it is not excluded that the process is competitive through the existence of a minor C$_2$H$_2$-loss photodissociation-channel, while H-loss dominates the photodissociation of [C-PAH]$^+$ complexes. Photodissociation experiments of PAHs, such as those of Marciniak et al.\ (2021), show that such minor photodissociation channels are often open with a low branching ratio of a few percent, in addition to  a main channel, such as H loss. This could be possible if, after the absorption of a UV photon, the C$_2$H$_2$-loss  rate remains non negligible compared to the PAH cooling rate; i.e., if N$_{\rm C}$ does not exceed a certain limit, that could be estimated through RRKM modelling if the transition state was known.  
Even with a very low branching ratio, br$_{\rm C2H2}$, this could lead to an  efficient C$_2$H$_2$ loss.

 Actually, after H loss leading to compounds Ph1 or Ph3 (Fig.\ \ref{fig:5-figure5}), one of the most probable processes is accretion of a new H leading back to C6 or C15. Therefore, carbon complexes of small PAHs, should spend a substantial fraction $\kappa$ of the time between two C$^+$ accretion in compounds similar to C6 or C15, so that the decay rate of [C-PAH] complexes through direct C$_2$H$_2$ loss is approximately $\kappa\times$ br$_{\rm C2H2}\times \gamma_{\rm dH}$, where $\gamma_{\rm dH}$ is the H-loss rate. Therefore, the chance of losing C$_2$H$_2$ before the next C$^+$ accretion may approach 100\% if $\kappa\times$ br$_{\rm C2H2}\times \gamma_{\rm dH}$ $>>$ $\gamma_{\rm C^+}$, i.e. $\kappa\times$ br$_{\rm C2H2}$ $>>$ $\gamma_{\rm C^+}$/$\gamma_{\rm dH}$ $\sim$ 10$^{-4}$, or 
 br$_{\rm C2H2}$\,$\ga$\,10$^{-3}$.

\subsection{ C$_2$H$_2$ loss through hydrogenation of [C-PAH]$^+$ complexes}
\label{app:C.3Hydrogenation}

The H-accretion rate  for coronene$^+$ was measured by Betts et al.\ (2006) as (1.4$\pm0.7)\times$  10$^{-10}$\,cm$^3s^{-1}$. 
It corresponds to the average rate $\gamma_{\rm aH}\sim$ \,0.1\,yr$^{-1}$, as assumed in Fig.\ \ref{fig:6A1-rates}. It was adopted by Andrews et al.\ (2016) for H accretion onto all PAH cations. It should be a lower value for H accretion onto radical [C-PAH]$^+$ complexes. 
Therefore, 
the fraction of superhydrogenated [C-PAH]$^+$ complexes of structure similar to H[C-coronene]$^+$ = C$_{25}$H$_{13}^+$ is about $\gamma_{\rm aH}$/$\gamma_{\rm UV}$\,=\,$\eta$ $\sim$\,1\%.

Marciniak et al.\ (2021) have shown that the branching ratios for the photodissociation of aliphatic  PAH derivative C$_{25}$H$_{13}^+$  are 70\% for C$_2$H$_2$ loss, 8\% for C$_2$H loss and 22\% for H loss. We may assume that these ratios are valid for interstellar photodissociation of similar aliphatic PAH derivatives for all values of N$_{\rm C}$\,$<$\,N$_{\rm Clim-aliph}$, where N$_{\rm Clim-aliph}$ is the limit for photo C$_2$H$_2$-loss of PAHs similar to aliphatic  PAH derivative C$_{25}$H$_{13}^+$, that depends on the unknown value of the activation energy of this process. Because of the efficiency of C$_2$H$_2$ loss from aliphatic PAH derivative C$_{25}$H$_{13}^+$, one may tentatively assume that  N$_{\rm Clim-aliph}$\,$\ga$\,50. 

Therefore, for such small PAHs, each time a [C-PAH]$^+$ complex is formed with an aliphatic structure similar to isomers C9 or C10 of [C-coronene]$^+$ (Fig.\ \ref{fig:3-figure3}), it is likely that it has a probability not much smaller than $\eta$ to form a hydrogenated aliphatic complex and to release a C$_2$H$_2$ molecule.
However, it is difficult to infer the overall rate of C$_2$H$_2$ loss from this process, because it depends on the complex cycle of H losses and re-accretions until the next formations of PAH-CH$_2^+$. Nevertheless, it might substantially contribute to the return of C$_2$H$_2$ to the interstellar gas from relatively small PAHs.

\section{PAH size distribution and accretion rates of C$^+$ and O}
\label{app:Dsize}

\subsection{Size distribution}
        \label{app:D.1sizedistribution}
It is impossible to derive a precise shape of the size distribution of interstellar PAHs because of our limited understanding of the processes which determine it. We therefore have to refer to empirical size distributions proposed on the basis of the ratios of the infrared spectral features (e.g. Desert et al.\ 1990; Draine \& Li 2007; Maragkoudakis et al.\ 2020; Knight et al.\ 2021).

As the precise form of this distribution does not matter for our discussions, we have adopted  the following simple model for the number of PAHs with N$_{\rm C}$ carbon atoms per unit volume, n$_{\rm PAH}$(N$_{\rm C}$), including: (i) a  N$_{\rm C}^{-1}$ dependence as Desert et al.\ (1990), that simplifies calculations; (ii) N$_{\rm C}$\,=\,25 and 150 for the lower and higher uncertain limits, respectively. As noted in Sect. \ref{sec:4.1summphotdiss},
the lower limit, N$_{\rm C}$\,=\,25, for the survival limit of PAHs in the diffuse ISM, is roughly inferred from Jochims et al.\ (1994) by including a dominant effect of recurrent fluorescence in PAH cooling (Lacinbala et al.\  2022; Bernard et al.\ 2023; Stockett et al.\ 2023, 2024).    
Such a lower limit for the diffuse ISM seems roughly consistent with the most recent estimates from JWST observations of the most shielded parts of Orion PDR regions by Chown et al.\ (2024). 
 The upper limit for the PAH size, N$_{\rm C}$\,=\,150, is again a compromise between early guesses, such as Desert et al.\ (1990) and Draine \& Li (2007), and the most recent estimates of the average PAH size  by Knight et al.\ (2021). 

In addition, we have normalized the PAH abundances, n$_{\rm PAH}$/n$_{\rm HT}$, so that the total carbon abundance locked in PAHs is X$_{\rm PAH}$\,=\,5\% of the total average interstellar abundance of carbon in the diffuse ISM, estimated as 2.4 $\times$ 10$^{-4}$ by Cardelli et al.\ (1996). Finally, we have assumed n$_{\rm HT}$\,=\,50\,cm$^{-3}$.

Therefore, we approximate the abundance of PAHs with N$_{\rm C}$ as 
\begin{equation}
        \label{eq:D1nPAH1}
{\rm n_{PAH}(N_{C})/n_{HT} \sim 10^{-7}/N_{C}} 
\end{equation} 
\begin{equation}
        \label{eq:D2nPAH2}
{\rm n_{PAH}(N_{C})\,\sim 5\times10^{-6}/N_{C}~{cm}^{-3}}
\end{equation} 
for ${\rm 25 <N_C<}$ 150. This yields a total  number of PAH molecules per unit volume n$_{\rm PAHT}$\,$\sim$\,10$^{-5}$\,cm$^{-3}$, and their total abundance $\chi _{\rm PAHT}$\,=\,n$_{\rm PAHT}$/n$_{\rm HT}$\,$\sim$\,2$\times10^{-7}$.

\bigskip

\subsection{Accretion rates of C$^+$ and O}
        \label{app:D.2CpOrates}

To model these rates in the diffuse ISM, we adopt the following fiducial values: total density of H nuclei, n$_{\rm HT}$\,=\,50\,cm$^{-3}$; carbon amount in PAHs, X$_{\rm PAH}$\,=\,5\% of its total interstellar amount, i.e.\ a number of C atoms in PAHs relative to hydrogen of 1.2 x 10$^{-5}$, yielding the above expressions for n$_{\rm PAH}$(N$_{\rm C}$)/n$_{\rm HT}$; half PAHs in cation and half in neutral form; average C$^+$ accretion rate on neutral PAHs, 2$\times$10$^{-9}$
\,cm$^3$s$^{-1}$ (Sect. \ref{sec:2.2CpPAH} and Canosa et al.\ 1995); C$^+$ abundance in the diffuse ISM, ${\rm n_{C+}/n_{HT}}$\,=\,1.4x10$^{-4}$;  O accretion rate on cation PAH$^+$s, 1.3$\times$10$^{-10}$\,cm$^3$s$^{-1}$ (Sect. \ref{sec:7CO} and Snow \& Bierbaum 2008); gaseous O abundance in the diffuse ISM, ${\rm n_{O}/n_{HT}}$\,=\,3.2x10$^{-4}$.

With these assumptions and the above expression for the abundance of PAHs,  one can estimate that, in the diffuse ISM, the characteristic times for the depletion of gaseous carbon and oxygen,  by accretion of  C$^+$ or O onto all PAHs (with 25\,$<$\,N$_{\rm C}$\,$<$\,150) are about 
\begin{equation}
        \label{eq:D3tauC}
{\rm \tau_C \sim 3 \times 10^6 ~yr} 
\end{equation}
\begin{equation}
        \label{eq:D4tauO}
{\rm \tau_O \sim 5 \times 10^7 ~yr} ,
\end{equation}
respectively. 
With the parameters assumed above for the diffuse ISM, this yields 
\begin{equation}
        \label{eq:D5RCp}
{\rm R_{C} = n_{C+}/\tau_C = 7x10^{-17} \times (X_{PAH}/5\%) \times (n_{HT} /50)^2 ~~ cm^{-3}s^{-1}},
\end{equation}
\begin{equation}
        \label{eq:D6RO}
{\rm R_{O} = n_O/\tau_O = 10^{-17} \times (X_{PAH}/5\%) \times (n_{HT} /50)^2 ~~ cm^{-3}s^{-1}}
\end{equation} 

for the absolute rates per unit volume for removing C$^+$ and O from the gas, respectively. 

\section{Simplified chemistry of C$_2$H$_2$, C$_2$H, and CO in the diffuse ISM}
\label{app:Echemistry}

Surprisingly, diffuse clouds are one of the areas of interstellar chemistry where several key questions remain not fully settled. The basic features of this chemistry were established in the seventies (e.g. Dalgarno \& Black 1976; Watson 1978), followed by detailed models (e.g. Black \& van Dishoeck 1986), mostly  combining the effects of cosmic rays, UV  photochemistry and ionic reactions, for accounting molecular abundances inferred from UV, optical and millimeter absorption spectroscopy. A key problem remained for long the high observed abundance of CH$^+$. It is now generally agreed that it results from high temperature reactions in turbulent shocks (e.g. Godard et al.\ 2014).  Reactions initiated by CH$^+$ may thus explain the high abundances observed for many species in the diffuse ISM, including C$_2$H (Gerin et al.\ 2010), HCO$^+$ and CO, as confirmed by recent developments (e.g. Gerin \& Liszt 2021). However, the whole problematics of the chemistry of the diffuse ISM remains a complex and active field of research (see e.g. references in Introduction of Rybarczyk et al.\ 2022, and Gerin et al.\ 2016).

Detailed discussions of such a chemistry is beyond the scope of this paper. Our goal in this appendix is just to sketch the role of a few key reactions for determining approximate estimates of the abundances of key molecules, C$_2$H$_2$, C$_2$H and CO, in order to check the importance of the outcome of accretion of C$^+$ and O onto PAHs for this chemistry. We use rate estimates for these reactions from the University of Manchester Institute of Science and Technology (UMIST) standard compilation of interstellar chemistry (McElroy et al.\ 2013).

\bigskip

\subsection{C$_2$H$_2$ and C$_2$H abundances from C$^+$ accretion onto PAHs}
\label{app:E.1abundances}

C$_2$H$_2$ and C$_2$H are mainly destroyed by photodissociation whose rate estimates are $\Gamma$$_{\rm C2H2}$\,=\,33$\times$10$^{-10}\times$ exp(-2.3\,A$_{\rm v}$)\,s$^{-1}$ and $\Gamma$$_{\rm C2H}$\,=\,5.2$\times$10$^{-10}\times$ exp(-2.3\,A$_{\rm v}$)\,s$^{-1}$, respectively, where A$_{\rm v}$ is the visible extinction (McElroy et al.\ 2013). These rates, $\sim$10$^{-9}$\,s$^{-1}$ and $\sim$10$^{-10}$\,s$^{-1}$, respectively, are one or two orders of magnitude faster than the rates of the next faster reactions with C$^+$, namely (McElroy et al.\ 2013): C$_2$H$_2$, k\,=\,1.8$\times$10$^{-9}$cm$^3$s$^{-1}$ $\rightarrow$ 1.3$\times$10$^{-11}$\,s$^{-1}$; C$_2$H, k\,=\,1.0$\times$10$^{-9}$cm$^3$s$^{-1}$ $\rightarrow$ 0.7$\times$10$^{-11}$\,s$^{-1}$. Calling R$_{\rm C2H2}$ the rate of injection of C$_2$H$_2$ per unit volume, the approximate number of C$_2$H$_2$ molecules per unit volume is n$_{\rm C2H2}$\,$\sim$\,R$_{\rm C2H2}$/$\Gamma$$_{\rm C2H2}$. Similarly, as R$_{\rm C2H2}$ is also the approximate rate of injection of C$_2$H, n$_{\rm C2H}$\,$\sim$\,R$_{\rm C2H2}$/$\Gamma$$_{\rm C2H}$.

 Assuming a typical value, A$_{\rm v}$\,=\,0.7, and that  C$_2$H$_2$ and  C$_2$H are mostly formed as outcomes of C$^+$ accretion onto PAHs, with a rate R$_{\rm  C2H2}$\,=\,0.25$\times$R$_{\rm C}$, given by Eqs. \ref{eq:2C2H2} and \ref{eq:D5RCp}, would yield for their abundances 
\begin{equation}
        \label{eq:E1chiC2H2}
{\rm n_{C2H2}/n_{HT} \sim 0.5x10^{-9}},
\end{equation}
\begin{equation}
        \label{eq:E2chiC2HnT}
{\rm n_{C2H}/n_{HT}~ \sim 3x10^{-9}}.
\end{equation}

 In many observational papers, molecular abundances are referred to the number of H$_2$ molecules per unit volume, n$_{\rm H2}$, rather than the total number of H nuclei, n$_{\rm HT}$, with n$_{\rm H2}$\,=\,0.5$\times$f$_{\rm H2}$$\times$n$_{\rm HT}$, where f$_{\rm H2}$ is the fraction of hydrogen in molecular form. This yields 

\begin{equation}
        \label{eq:E3chiC2HH2}
{\rm n_{C2H}/n_{H2} \sim 0.7x10^{-8}\times f_{H2}^{-1}}
\end{equation}
or, for a typical value f$_{\rm H2}$\,=\,0.4, 
${\rm n_{C2H}/n_{H2} \sim 2x10^{-8}}$.

\bigskip

\subsection{CO injection from O accretion onto PAHs}
\label{app:E.2CO}
Assuming that each O accretion onto PAHs leads to the formation of a CO molecule yields that the CO formation rate equals R$_{\rm O}$ (Eq. \ref{eq:D6RO}): 
\begin{equation}
        \label{eq:E4RCO1}
{\rm R_{CO-PAH} ~\sim 10^{-17}~~ cm^{-3}s^{-1}}.
\end{equation}
In classical models of chemistry of diffuse clouds, CO is formed mainly from electron recombination of HCO$^+$ + e$^-$  $\rightarrow$ CO + H (e.g.\, Gerin \& Liszt 2021). The order of magnitude of the actual rate of CO formation,  R$_{\rm CO}$, may be inferred from the observed abundance of HCO$^+$, ${\rm n_{HCO+}/n_{H2}}\,=\,3\times 10^{-9}$ (Gerin \& Liszt 2021). Assuming a recombination rate for HCO$^+$, ${\rm \alpha \,=\,1.45\times 10^{-5}/T^{0.69}\,cm^3s^{-1}}$ (Hamberg et al.\ 2014), n${\rm _e=n_{C+}}$, T\,=\,100\,K, and n${\rm _{H2}/n_{HT}}$\,=\,0.2 yields
\begin{equation}
        \label{eq:E5RCO2}
{\rm R_{CO} ~= 1.3x10^{-16} ~~ cm^{-3}s^{-1}}.
\end{equation}
This value is an order of magnitude greater than the rate of formation of CO from oxygenation of  PAHs (Eq. \ref{eq:E4RCO1}), which, therefore, contributes marginally to the formation of CO in diffuse clouds.

\section{Additional figures}
\begin{figure*}[htbp]
        \begin{center}
                \includegraphics[scale=1.05, angle=0]{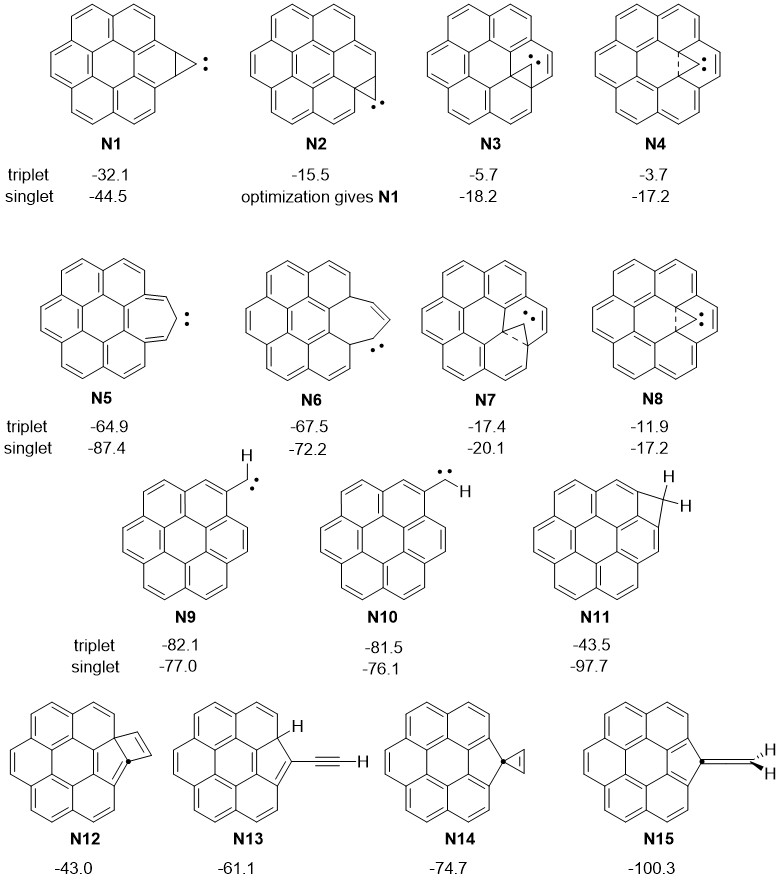}
                \caption{Energies (kcal/mol) of various isomers of neutral [C-coronene] complexes, C$_{25}$H$_{12}$, as computed at B3LYP/6-311+G** + ZPVE (see Sections \ref{sec:3.3Ccoronene} and \ref{sec:3.1method}).}
                \label{fig:4-figure4}
        \end{center}
\end{figure*}

\begin{figure*}[htbp]
        \begin{center}
                \includegraphics[scale=0.9, angle=0]{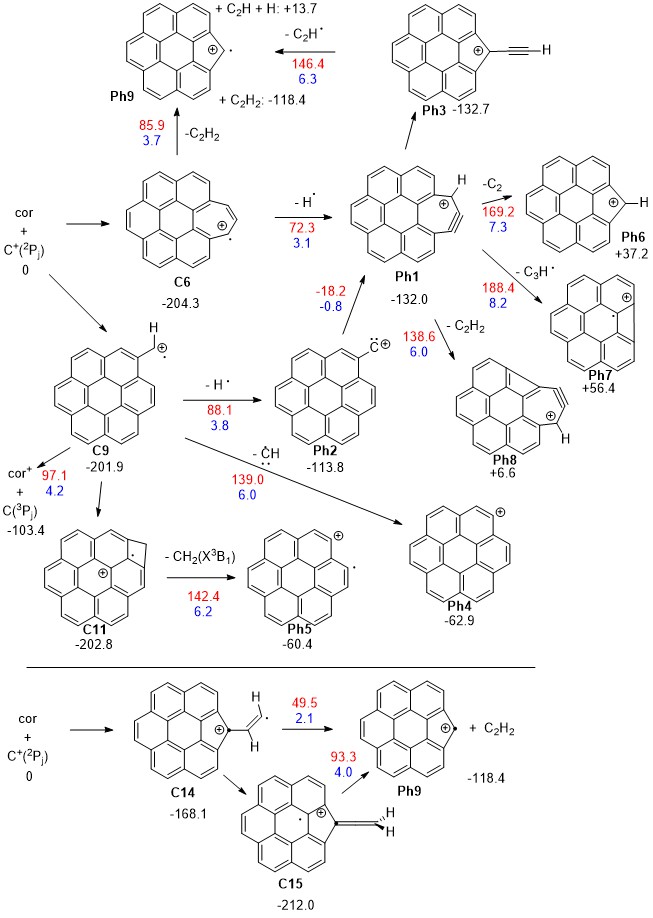}
                \caption{Energies [kcal/mol, vs coronene + C($^2$P$_{\rm j}^+$)] of different possible structures into which the most stable isomers of C-coronene$^+$ cation, C$_{25}$H$_{12}^+$,  C6 (with a 7C ring), C14, and C15 (Fig.\ \ref{fig:3-figure3}), could be photofragmented (Sections  \ref{sec:3.4Ephotodiss} and  \ref{sec:4.2processingsmall}). The reaction energies for the various fragmentation processes are given in kcal/mol (red) and eV (blue).}
                \label{fig:5-figure5}
        \end{center}
\end{figure*}

Figure\ \ref{fig:4-figure4} displays the computed energies of various isomers of neutral [C-coronene] complexes, C$_{25}$H$_{12}$ (see Sect. \ref{sec:3.3Ccoronene}).
Figure\ \ref{fig:5-figure5} displays the energies  of different possible structures into which the most stable isomers of C-coronene$^+$ cation, C$_{25}$H$_{12}^+$,  C6 (with a 7C ring), C14, and C15, could be photofragmented (see Sect. \ref{sec:3.4Ephotodiss}).

\end{document}